\documentclass[trackchanges,twocolumn]{aastex701}
\usepackage{xspace}
\usepackage{amssymb}
\usepackage{stix}
\usepackage{rotating}       
\usepackage{booktabs}       
\usepackage{threeparttable} 

\newcommand{\jplephemdate}{\mbox{2025-07-28}\xspace}
\newcommand{\discovery}{\mbox{2025-07-01}\xspace}

\newcommand{\mpclcdate}{\mbox{2025-09-15}\xspace}
\newcommand{\tessfirst}{\mbox{2025-05-07}\xspace}
\newcommand{\tesslast}{\mbox{2025-06-02}\xspace}
\newcommand{\cname}{3I/ATLAS\xspace}
\newcommand{\fficubedr}{\mbox{2025-06-13}\xspace}

\graphicspath{{./}{figures/}}

\begin{document}

\title{Pre-discovery TESS Observations of Interstellar Object 3I/ATLAS}

\author[orcid=0000-0002-7395-4935]{Jorge Martinez-Palomera}
\affiliation{University of Maryland, Baltimore County, 1000 Hilltop Circle, Baltimore, Maryland, United States}
\affiliation{NASA Goddard Space Flight Center, 8800 Greenbelt Road, Greenbelt, MD 20771, USA}
\email[show]{jorge.i.martinezpalomera@nasa.gov, jorgemp@umbc.edu}  

\author[orcid=0000-0002-2830-9064]{Amy Tuson}
\affiliation{University of Maryland, Baltimore County, 1000 Hilltop Circle, Baltimore, Maryland, United States}
\affiliation{NASA Goddard Space Flight Center, 8800 Greenbelt Road, Greenbelt, MD 20771, USA}
\email{}

\author[orcid=0000-0002-3385-8391]{Christina Hedges}
\affiliation{University of Maryland, Baltimore County, 1000 Hilltop Circle, Baltimore, Maryland, United States}
\affiliation{NASA Goddard Space Flight Center, 8800 Greenbelt Road, Greenbelt, MD 20771, USA}
\email{}

\author[orcid=0000-0003-4206-5649]{Jessie Dotson}
\affiliation{NASA Ames Research Center, Moffett Field, CA 94035, USA}
\email{}

\author[orcid=0000-0001-7139-2724]{Thomas Barclay}
\affiliation{NASA Goddard Space Flight Center, 8800 Greenbelt Road, Greenbelt, MD 20771, USA}
\email{}

\author[orcid=0000-0003-0501-2636]{Brian Powell}
\affiliation{NASA Goddard Space Flight Center, 8800 Greenbelt Road, Greenbelt, MD 20771, USA}
\email{}

\begin{abstract}

\cname, also known as C/2025 N1 (ATLAS), is the third known macroscopic interstellar object to pass through our Solar System. We report serendipitous Transiting Exoplanet Survey Satellite (TESS) observations of \cname taken between \tessfirst and \tesslast, 55 days prior to the discovery date (\discovery). We retrieve the TESS pixel data, perform a robust background correction and use a data-driven approach to compute the object's position on the TESS detectors. We find a consistent offset between the target's observed and predicted positions which is dominated by uncertainty in the TESS World Coordinate System (WCS) rather than ephemeris errors. \cname is too faint to be detected in the individual 200\,second TESS integrations, so we stack images to improve detectability. We perform aperture and Pixel Response Function (PRF) photometry on the stacked images to create two light curves. Each light curve consists of 15 measurements with $\text{SNR}>3$, collected across two different TESS cameras during the 26\,days that the object was observed. The PRF light curve, which is more robust against image noise, in the TESS bandpass shows  a gradual increase in brightness from $T_{\text{mag}}=20.9\pm0.29$ to $19.57\pm0.15$. This is expected as \cname approaches the inner Solar System. Its absolute magnitude decreases from $H_{V}=14.3\pm0.4$ to $13.7\pm0.3$ and shows signs of faint activity consistent with other observations. This paper highlights the power of using TESS for Solar System science; by increasing the number of pre-discovery observations, in an otherwise sparsely populated region of the light curve, the long-term behavior of \cname can be investigated.

\end{abstract}


\section{Introduction}\label{sec:intro}

The interstellar object \cname, or C/2025 N1 (ATLAS), was discovered by the Asteroid Terrestrial-impact Last Alert System \citep[ATLAS;][]{2018PASP..130f4505T} in Chile on \discovery \citep{2025MPEC....N...12D}. This is only the third confirmed macroscopic interstellar object, following the discoveries of 1I/`Oumuamua in 2017 \citep{2017CBET.4450....1W} and 2I/Borisov in 2019 \citep{2019CBET.4666}. One possible explanation for their origin is that these objects form in protoplanetary disks around other stars before being ejected due to events such as close stellar flybys and planetary scattering \citep[e.g.,][]{2023ARA&A..61..197J,2023arXiv230317980F}. Studying their composition and dynamics therefore enables us to better understand planet formation processes in other star systems and infer possible ejection mechanisms.

Following the discovery of \cname, numerous follow-up observations have been coordinated and several pre-discovery detections have been made using archival data. The first characterization of the object by \citet{2025arXiv250702757S} revealed an orbital eccentricity $e \sim 6.1$ and a velocity at infinity $V_{\infty} \sim 58\,\text{km\,s}^{-1}$, as well as signs of cometary activity and an upper limit for the object's radius ($\sim10$\,km). Observations from the Vera C. Rubin Observatory were used to derive a nucleus radius of $r_n\sim5.6$\,km \citep{2025arXiv250713409C}, while observations from the Hubble Space Telescope limit the effective radius of the nucleus to $r_n\le2.8$ km \citep{2025ApJ...990L...2J}. A low-amplitude rotation signal with period $16.79$\,hrs was reported by \citet{2025arXiv250712922D} and a similar period of $16.16 \pm 0.01$ h was detected by \citet{2025arXiv250800808S}. However, several other studies have not detected this photometric rotation signal \citep{2025arXiv250702757S,2025arXiv250713409C,2025arXiv250712234K}.
Targeted pre-perihelion observations have discovered dust composition similar to D-type asteroids, a $\text{CO}_2$ dominated coma, and the presence of $\text{H}_2\text{O}$, CO, OCS, water ice and dust \citep{2025arXiv250714916Y,2025arXiv250818209C}.

In this paper, we present Transiting Exoplanet Survey Satellite \citep[TESS;][]{2015JATIS...1a4003R} observations of \cname taken between \tessfirst and \tesslast, 55 days before the object was discovered. This is the first time TESS has observed an interstellar object. This paper is organized as follows. In Section~\ref{sec:data}, we present the TESS observations and data analysis, in Section~\ref{sec:results} we present our results and in Section~\ref{sec:disc_conc} we present a discussion of our findings.

\section{Observations and Data Analysis}\label{sec:data}

TESS is an all-sky photometric survey that began science operations in 2018 and orbits the Earth approximately once every two weeks in a high altitude, highly eccentric orbit. While TESS was launched with the primary goal of discovering new transiting exoplanets, it has also contributed significantly to planetary science \citep[e.g.,][]{2019ApJ...886L..24F,2020ApJS..247...26P,2021PASP..133a4503W,2025PASP..137d4401T}. TESS has four identical cameras, each of which focuses light onto an array of four charge-coupled devices (CCDs), with a large combined field of view of $24^{\circ}\times96^{\circ}$. It observes the sky in a series of approximately 27\,day long observing sectors, providing almost continuous photometry with an observing cadence that has decreased throughout the mission -- 30\,minutes during the two-year primary mission, 10\,minutes during the first extended mission and now 200\,seconds. The TESS mission delivers Full Frame Images (FFIs) of the entire instrument field of view from each sector and these are downlinked from the spacecraft at the end of each week. These FFIs are a valuable resource for time-series photometry.

We used the \texttt{tess-ephem}\footnote{\texttt{tess-ephem} is an open-source Python package that checks the observability of Solar System objects with TESS. \url{https://github.com/SSDataLab/tess-ephem}} software to check if and when \cname was observed by TESS. It was observed during sector 92, in the second extended mission, between \tessfirst and \tesslast. \cname was observed by TESS for a total of $\sim26$\,days -- 10.9\,days (4,727 frames) in camera 2 CCD 3 and then 14.7\,days (6,364 frames) in camera 1 CCD 2. During this time, the heliocentric distance of \cname decreased from 6.35 to 5.46 AU.

In the following sections, we outline our data analysis methods. Briefly, we retrieve the relevant TESS FFI data (Section \ref{subsec:data_red}), model and subtract the background flux (Section \ref{subsec:bkg_corr}), compute the target's offset from its expected position using data-driven methods (Section \ref{subsec:ephem_corr}), stack individual frames to increase the significance of the detection (Section \ref{subsec:stacking}) and extract the target's light curve using aperture and Pixel Response Function (PRF) photometry (Section \ref{subsec:phot}).

\subsection{TESS Data}\label{subsec:data_red}

We used version 1.2.6 of the \texttt{tess-asteroids} package\footnote{\texttt{tess-asteroids} is an open-source Python package to create image cutouts and light curves for Solar System objects observed by TESS. \url{https://github.com/altuson/tess-asteroids}} \citep{tuson_2025_16332750} to efficiently access and process the TESS pixel data. With \texttt{tess-asteroids}, we queried the JPL Horizons\footnote{\url{https://ssd.jpl.nasa.gov/horizons}} system to retrieve the orbital elements and ephemeris of \cname during the sector 92 observing window. It then calculates the expected position of the target on the TESS detectors using \texttt{tesswcs}\footnote{\texttt{tesswcs} is an open-source Python package that accesses the TESS World Coordinate System (WCS) solutions. This enables conversion between Right Ascension and Declination into pixel coordinates in TESS cameras. \url{https://www.github.com/tessgi/tesswcs}} and uses \texttt{tesscube}\footnote{\texttt{tesscube} is an open-source Python package that enables efficient access to TESS FFI cubes. \url{https://github.com/tessgi/tesscube}} to retrieve the time-series data for pixels within a square region ($21\times21$\,pixels) centered on the target position at the mid-time of each TESS frame. \texttt{tesscube} accesses the Science Processing Operations Center \citep[SPOC;][]{spoc} FFI cubes avilable on the Mikulski Archive for Space Telescopes's (MAST's) Amazon Web Services (AWS) cloud storage environment. These FFI cubes contain the calibrated images. The complete set of sector 92 FFIs were made available on \fficubedr in data release 126\footnote{\url{https://archive.stsci.edu/missions/tess/doc/tess_drn/tess_sector_92_drn126_v01.pdf}} \citep[MAST Dataset][]{mast_tess_ffi}. The result from \texttt{tess-asteroids} is a Target Pixel File (TPF) -- an image cutout that tracks \cname as it moves across the TESS FFI.

We used the data quality flags in the SPOC FFI headers to remove all cadences that were identified as low quality, including times where strong background light was present and when the spacecraft lost fine pointing control. After removing low quality data, 4,505 ($95.3\%$) and 5,206 ($81.8\%$) good quality frames remained for camera 2 CCD 3 and camera 1 CCD 2, respectively. This is the data that is used in the following sections. Table~\ref{tab:frame_counts} summarizes the number count of selected frames and pixels.

\begin{deluxetable}{lrr}
\tablewidth{0pt}
\tablecaption{TESS Sector 92 Observation Summary  \label{tab:frame_counts}}
\tablehead{
\colhead{} & \colhead{Camera 2 CCD 3} & \colhead{Camera 1 CCD 2}
}
\startdata
Observing baseline (d) & 10.9 & 14.7\\
Total $N_{\text{frames}}$ & 4,727 & 6,364 \\
Good quality $N_{\text{frames}}$ & 4,505 & 5,206 \\
Strong scattered light$^{\dag}$ & 676 & 781 \\
Total $N_{\text{pix}} \times N_{\text{frames}}$ & 35,599,037 & 88,332,320 \\
Good quality $N_{\text{pix}} \times N_{\text{frames}}$ & 33,927,155 & 72,259,280 \\
\enddata
\tablecomments{$N_{\text{frames}}$ is the number of frames. $N_{\text{pix}}$ is the number of unique pixels. $^{\dag}$ This is the number of good quality frames that have strong scattered light (see Section~\ref{subsec:bkg_corr}).}
\end{deluxetable}


\subsection{Background Correction} \label{subsec:bkg_corr}

TESS FFI observations are strongly affected by a systematic background signal that varies with detector position and time. The largest contribution to the background signal is from the Earth and Moon, which create background scattered light across all cameras and CCDs. This background is strongest when the Earth or Moon are close to the TESS boresight during the orbit, which can happen up to twice in a sector. During sector 92, 676 and 781 cadences show significant scattered light signal (median flux in the top 15 percentile) for camera 2 CCD 3 and camera 1 CCD 2, respectively.

For the purpose of our analysis, we have parameterized the background signal in TESS as consisting of two components; a broad, time dependent background due to Earth and Moon scattered light reaching the detector, and a background of stars behind \cname. Both of these must be accounted for to obtain a high-quality light curve for a faint target like \cname.

To model the background scattered light, we first have to mask the target and background stars. We use the PRF model of TESS to mask the target. The PRF can be evaluated at any position on the TESS detector and is unique per camera and per CCD. We use the Python package \texttt{lkprf}\footnote{\texttt{lkprf} is an open-source Python package that enables evaluation of the TESS PRF model, collected from engineering data, at any position on the detectors. \url{https://www.github.com/lightkurve/lkprf}} to evaluate the PRF at each position that \cname lands on the detector. For a moving object, the model is the result of the PRF convolved with the object's path on the detector during the exposure. We model this as the sum of PRFs evaluated on the interpolated target position every minute. This is sufficient to capture the expected movement of \cname in a single 200\,second exposure, since its average speed is $0.6\,\text{pix\,arcmin}^{-1}$. The resulting PRF model gives the expected fraction of the target's total flux that falls on each pixel. We mask pixels which contain more than 1\% of the target’s total flux, resulting in an average of 15 pixels being masked per cadence.
Using all frames to obtain a time-averaged image, we mask stars by identifying pixels with a flux value larger than 1.1 times the median flux in the average image and a flux gradient value larger than 3 times the median gradients of the average image with respect to both column and row. By using the image gradient in combination with the flux threshold, we are ensuring the mask only includes stars and not extended structures in the background. The mask removes $\sim\!30\%$ of the pixel data and is sensitive to stars with $T_{\text{mag}}\lesssim 16$, where $T_{\text{mag}}$ is the magnitude in the TESS bandpass. For the remaining background pixels, we use Singular Value Decomposition \citep[SVD;][]{2009arXiv0909.4061H} to obtain the top 5 time-series components of the data. Using SVD, we model the data as

\begin{equation}
    A'=U \Sigma V'^T\ ,
\end{equation}

where $A'$ is the background model for unmasked pixels, indicated by the $'$ symbol. $U$ is a set of 5 column vectors with the most prominent time-series variability in the dataset. $V'$ describes the spatial layout of each of these components for all unmasked pixels. 
In order to ``infill'' the pixels that were masked from this fit, we fit a two-dimensional B-spline in row and column position of the unmasked pixels ($\mathbf{r}'$ and $\mathbf{c}'$) to $V'$. We use a spline with knots spaced every 25 pixels to capture broad spatial structure. Fitting a spatial model to $V'$ enables us to infill pixels which were masked out, so that we can find $V$ and calculate

\begin{equation}
    A=U \Sigma V^T\ .
\end{equation}

$A$ is a model of all pixel time-series, which contains the principal components of the scattered light variability. This process of infilling is only possible because the most prominent components of the background in TESS have a strong spatial dependency. Subtracting the model $A$ from the data removes the scattered background light. 

To model the stars behind \cname, we fit each pixel time-series separately. We mask out times within each pixel time-series which may contain flux from \cname, using the same target mask described earlier in this section. We also mask out any significant outliers in the pixel time-series. We then a) subtract the scattered light background model from each pixel and b) fit a B-spline in time to the pixel time-series. This B-spline has knots spaced every 6\,hours, necessary to capture relatively fast stellar variability. We opted to use B-spline functions, over others such as polynomials, to enable controlled flexibility of a smooth model. This allows us to model the flux of any star in the background, while removing stellar variability and flux changes from systematics such as differential velocity aberration. This B-spline model is the final model for the star field.

\begin{figure*}
    \centering
    \includegraphics[width=0.48\linewidth]{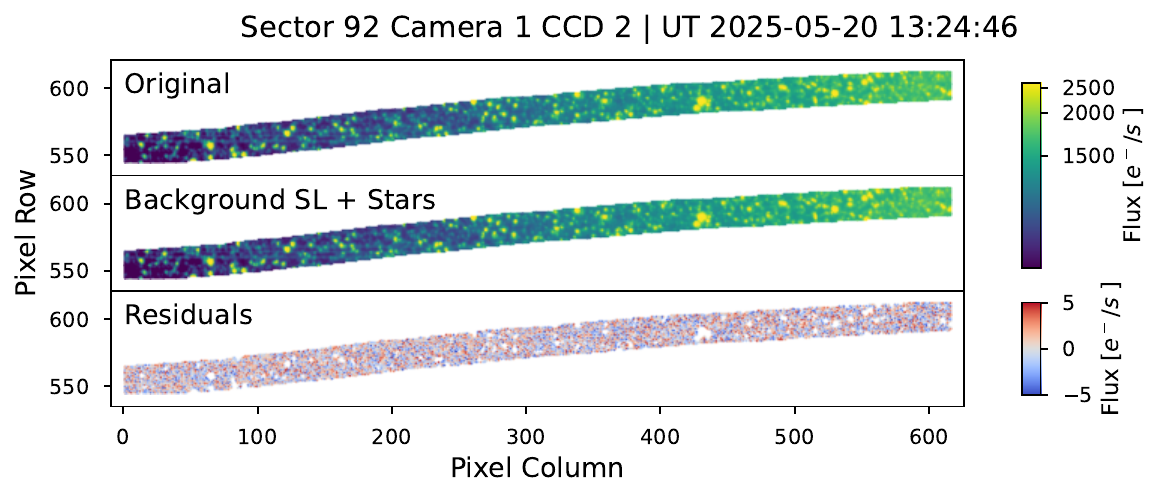}
    \includegraphics[width=0.48\linewidth]{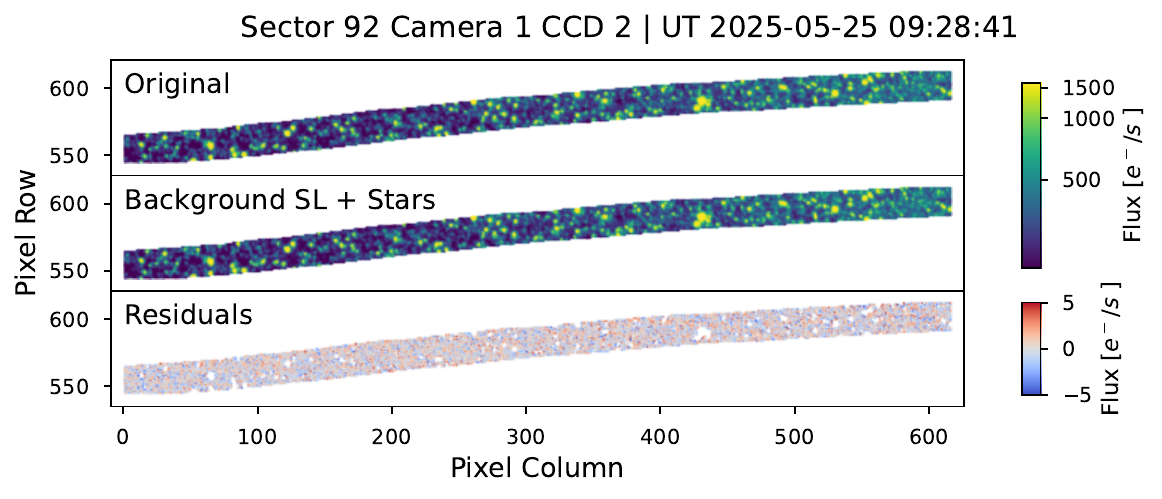}
    \caption{Background modeling, as described in Section~\ref{subsec:bkg_corr}. The left/right panel shows an example of a cadence with/without strong scattered light (SL). We show the raw flux (top), our background model that accounts for scattered light and stars (middle) and the residuals (bottom). The ``holes'' in the residuals are pixels that were identified as outliers and rejected.}
    \label{fig:bkg_model}
\end{figure*}

After subtracting both the scattered light model and the star model from the TPF, we are left with the background corrected pixel data.
We finally reject pixels with residual background signal (e.g., around bright stars and overly crowded regions). If a pixel has any outliers in its time-series, identified by $3\sigma$ clipping, it gets rejected at all times. This process removes $\sim10\%$ of the pixels. 
Figure~\ref{fig:bkg_model} shows the results of our background modeling. It includes examples of the raw flux, scattered light plus star model and residuals for cadences with and without strong scattered light. The rejected pixels show up as holes in the residual maps.

Our final background corrected flux has a mean and standard deviation of $0.006 \pm 1.6\, e^{-}/s$, and a median and median absolute deviation of $0.005 \pm 1.0\, e^{-}/s$. This corrected set of pixel time-series is used for the following analyses.

\subsection{Position Correction Model}\label{subsec:ephem_corr}

The position of \cname on the TESS detector as a function of time is dictated by i) the astrophysical ephemeris of the object and ii) the systematic distortions of the TESS instrument. There are uncertainties in both of these components, but ensuring the position of the target is accurate is essential for image stacking (see Section~\ref{subsec:stacking}) and photometry (see Section~\ref{subsec:phot}). We therefore use the TESS data to refine the positions of \cname in pixel space, retrieving the best fit model of the target's position over time.

\cname was faint when observed by TESS. The predicted visual magnitude from JPL Horizons was $V_{\text{mag}}\sim18.9-19.7$, equivalent to $T_{\text{mag}}\sim18.2-18.9$ using the conversion from \citet{2021PSJ.....2..236F} ($T_{\text{mag}} = V_{\text{mag}} - 0.8$). \cname is therefore not detectable in an individual 200\,second TESS exposure. As such, fitting the ephemeris of the object is confounded by noise. Rather than fitting the ephemeris from the TESS data alone, we instead assume that the reported ephemeris from JPL Horizons (solution accessed on \jplephemdate) is broadly correct and fit for a small perturbation in the ephemeris to best match the TESS data. We use the background corrected data (described in Section~\ref{subsec:bkg_corr}) for our modeling. \cname is approximately centered in each frame, with sub-pixel deviations. The true center of the object within any cutout is given by ($r_t$, $c_t$), the row and column position at time $t$.

We firstly assume that \cname is well modeled by the TESS PRF, as described in Section~\ref{subsec:bkg_corr}, multiplied by some unknown amplitude in every frame. The PRF for a given camera and CCD at time $t$ is given by 

\begin{equation}
p(r_t, c_t)\ ,
\end{equation}

which is an approximately $21\times21$ pixel image at time $t$. The values of this image are between 0 and 1, and will sum to 1. This image contains the PRF of TESS, evaluated at the predicted target position. We assume the data is fit well by the model

\begin{equation}
    y_t = a_t  p(r_t, c_t)\ ,
\end{equation}

where $a_t$ is the flux amplitude of the target at time $t$, assuming that $p$ is evaluated at the accurate ephemeris. This is a simple linear model and it is therefore easy to find $a_t$ using the data. However, because of systematics within the TESS detector and uncertainties on the true ephemeris, we do not know that $p$ is accurate. We would like to retrieve the amplitude of the object in TESS data and also find the best fitting position of the object ($r_t$, $c_t$). In order to preserve this as a linear model and search for only a small perturbation we will use Taylor expansion. Our model is given as
\begin{equation}
y_t = a_t  p(r_t + \Delta r_t, c_t + \Delta c_t)\ ,
\end{equation}

where $\Delta r_t$ and $\Delta c_t$ represent small perturbations in the row and column positions of the target at time $t$. Using Taylor expansion we approximate

\begin{equation}
    y_t \approx a_t p(r_t, c_t)  + a\Delta r_t\frac{\partial p(r_t, c_t)}{\partial r} + a\Delta c_t \frac{\partial p(r_t, c_t)}{\partial c}\ ,
\end{equation}

where $\frac{\partial p(r_t, c_t)}{\partial r}$ and $\frac{\partial p(r_t, c_t)}{\partial c}$ are the partial derivatives of $p(r_t, c_t)$ with respect to row and column, respectively. We estimate these gradients numerically. This model now represents a linear model for the flux of the target at any given time $t$, allowing for a small (sub-pixel) shift in row and/or column. Using this model we can fit to find $a_t$, $\Delta r_t $ and $\Delta c_t$ for each time $t$. 

Following our assumption that the \cname ephemeris is close to correct and therefore the corrections are small, we model the perturbation as

\begin{equation}
\label{eq:perturbation1}
 \mathbf{\Delta r} = \sum_{i=0}^nw_{i}\mathbf{t}^i
\end{equation}
\begin{equation}
\label{eq:perturbation2}
 \mathbf{\Delta c} = \sum_{i=0}^nv_{i}\mathbf{t}^i
\end{equation}
\begin{equation}
\label{eq:amp1}
 \mathbf{a} = \sum_{i=0}^mz_{j}\mathbf{t}^j\ ,
\end{equation}

where $\mathbf{\Delta r}$ and $\mathbf{\Delta c}$ are vectors of the perturbation in row and column as a function of the vector of all times $\mathbf{t}$ and $\mathbf{a}$ is the amplitude of the object in TESS data as a function of $\mathbf{t}$. These are simple polynomial models with order $n$ for the row and column perturbations and $m$ for the amplitude model. We select low-order polynomials ($n=m=3$) as a simple, linear model which enables us to encode an amplitude or a perturbation that slowly changes over time, with the value in a given time being strongly predicted by neighboring times. The data at all times, $y$, is then given by the model

\begin{equation}
\label{eq:model}
    y \approx 
    \mathbf{a}  p(\mathbf{r}, \mathbf{c})
    + \mathbf{a} \mathbf{\Delta r}\frac{\partial p(\mathbf{r}, \mathbf{c})}{\partial r} +
    \mathbf{a} \mathbf{\Delta c} \frac{\partial p(\mathbf{r}, \mathbf{c})}{\partial c}\ .
\end{equation}

By substituting Equations~\ref{eq:perturbation1}, ~\ref{eq:perturbation2} and ~\ref{eq:amp1} into Equation~\ref{eq:model} we can write the model as a set of linear equations. Because this model is linear, we are able to simply and efficiently fit this model to the data to retrieve the weights $w_i$, $v_i$, and $z_j$ without sampling. These weights can then be used in Equations~\ref{eq:perturbation1}, ~\ref{eq:perturbation2} and ~\ref{eq:amp1} to find the vectors $\mathbf{\Delta r}$ and $\mathbf{\Delta c}$ which describe the perturbation we must apply to best fit the ephemeris to the data at each time step. 

We fit the model to all pixels close to the expected position of \cname, where the pixel quality from \texttt{tess-asteroids} indicates no saturated pixels (flux value less than $10^5 e^{-}/s$), and at times where the SPOC cadence quality flags indicate no data issues. We fit the data resulting from different TESS CCDs separately, since each CCD has a unique PRF model. We perform this fit in two steps, first retrieving an estimate of the amplitude, $\mathbf{a}$, with no perturbation to the ephemeris, and then refining the amplitude and position. The results of this fit can be evaluated at any time. 

We evaluate the model at the times corresponding to our data and convert the resulting pixel coordinates into world coordinates using \texttt{tesswcs}. This package uses the World Coordinate System (WCS) from the relevant TESS sector, camera, CCD and updates the simple imaging polynomial (SIP) distortion coefficients with the mission lifetime average for that camera/CCD. Our results are presented in Table~\ref{tab:lc}, in columns Right Ascension (RA) and Declination (Dec), and in Section~\ref{subsec:res_astro}.

\subsection{Image Stacking}\label{subsec:stacking}

\begin{figure}[h!]
    \centering
    \includegraphics[width=.99\linewidth]{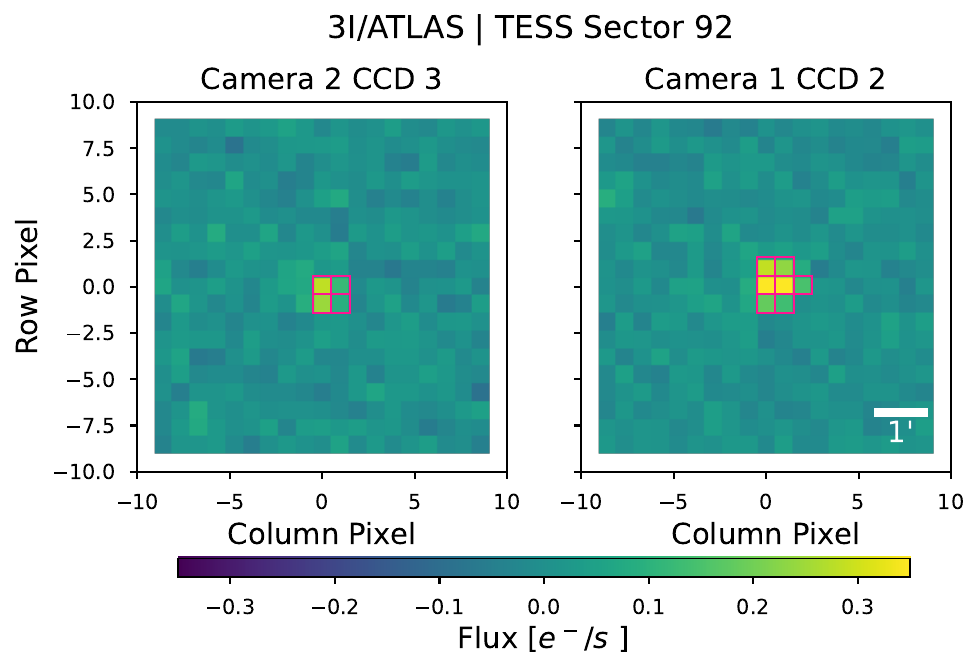}
    \caption{Full image stack of \cname, using all frames available from TESS sector 92 after quality filtering. Data from camera 2 CCD 3 (left) and camera 1 CCD 2 (right) are stacked separately to preserve instrument characteristics. The pixels with $\text{SNR}>3$ are highlighted in red; these are used for aperture photometry (see Section~\ref{subsec:phot}).}
    \label{fig:full_stack}
\end{figure}

The expected brightness of \cname during the TESS observations is below the detection limit of the 200\,second exposures ($T_{\text{mag}}\sim17.5$), meaning that detection in the single frames would be dominated by background noise.
Therefore, we performed image stacking to improve the Signal-to-Noise Ratio (SNR) and increase the detection limit to fainter magnitudes.
We used the corrected positions computed in Section~\ref{subsec:ephem_corr} to shift each frame and compensate for the movement of the target.
The image is interpolated with a third-order spline function to the shifted pixel grid.
We then calculated the median of each pixel to compute the stacked image.
Figure~\ref{fig:full_stack} shows the stacked images using all available frames for each camera. The pixels highlighted in red have $\text{SNR}>3$.

\begin{figure*}[h!]
    \centering
    \includegraphics[width=.99\linewidth]{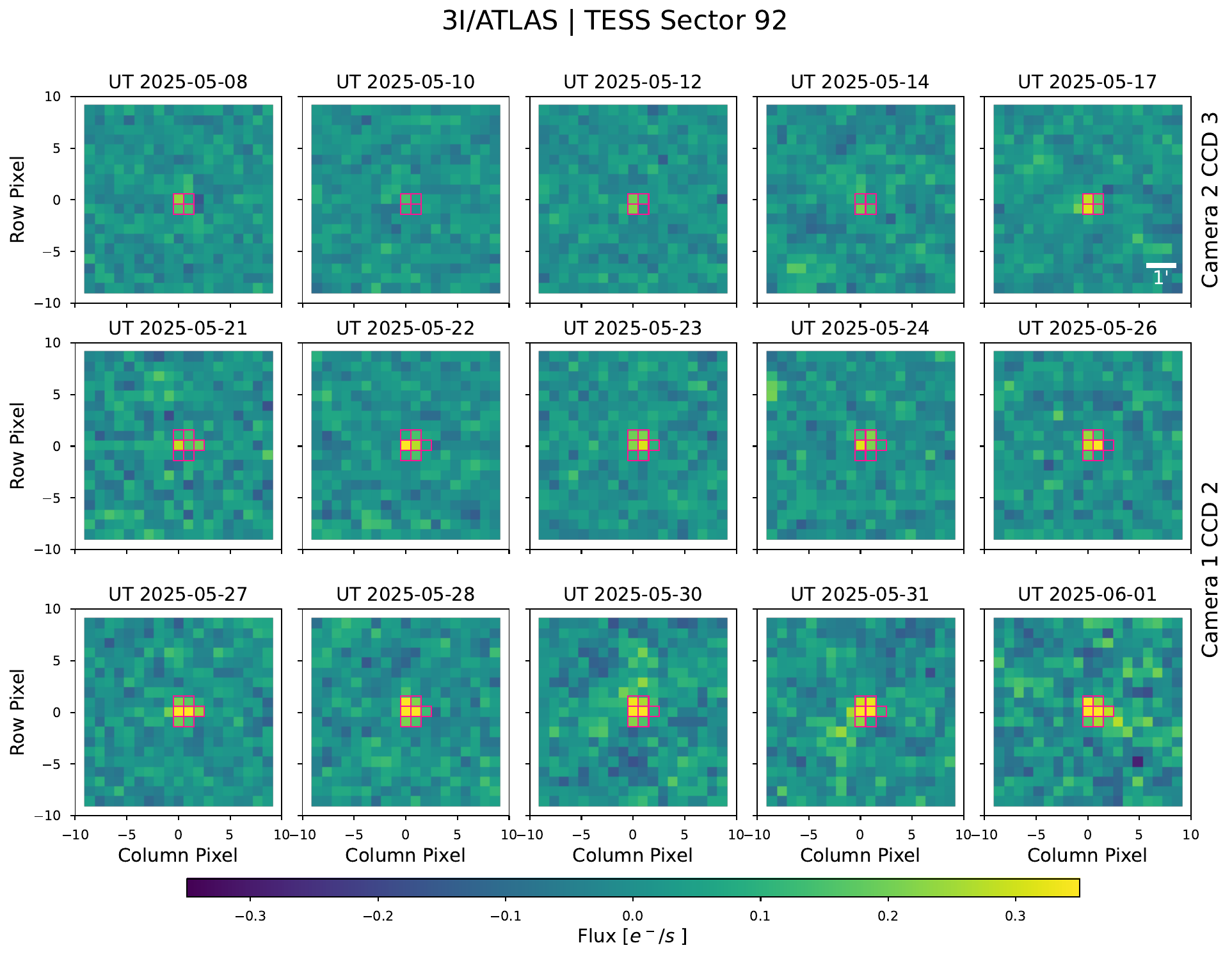}
    \caption{Stacked images of \cname from TESS sector 92. We stacked 901 ($\sim 2.1$\,days) and 520 ($\sim 1.2$\,days) frames from camera 2 CCD 3 (first row) and camera 1 CCD 2 (second and third row), respectively. The aperture masks used for photometry are plotted in red; the masks are different for each camera to account for instrument characteristics. Due to the stacking procedure, the row and column pixel values do not represent the location in the original TESS FFI and are only shown for reference. The images are aligned so that the target is always centered. 
    See Section~\ref{subsec:stacking} for details on the stacking procedure and Section~\ref{subsec:phot} for details on the photometry.}
    \label{fig:image_stack}
\end{figure*}

In order to assess the number of frames necessary in a stack to achieve a $3\sigma$ detection, we cumulatively stacked frames until at least one pixel within a 3-pixel radius from the center of the image reached $\text{SNR}>3$.
We found that for camera 2 CCD 3, at the beginning of sector 92 when \cname was at its faintest, we need to stack $N_{\text{frames}}\sim 900 \sim 2.1$\,days.
The number of stacked frames for camera 1 CCD 2 is $N_{\text{frames}}\sim 520 \sim 1.2$\,days.
During these time windows, \cname traveled 0.07 and 0.04 AU, respectively.
The smaller number of frames needed in the second CCD is due to the increase in brightness of \cname.
Because of the different number of available frames in each camera/CCD and the different number of frames that were stacked, camera 2 CCD 3 has five final stacked images and camera 1 CCD 2 has ten.
Figure~\ref{fig:image_stack} shows these 15 stacked images.
A clear signal above the background noise is visible in the center of most of the images, but it is more noticeable in camera 1 CCD 2.

\subsection{Photometry} \label{subsec:phot}

With our stacked images, we extract a light curve of \cname using aperture photometry. The aperture mask was defined by stacking all available frames from each camera/CCD separately and selecting all pixels within three pixels of the image center with $\text{SNR}>3$ (see Figure~\ref{fig:full_stack}). 
This minimized the number of background pixels included in the aperture and created a unique mask for each CCD, necessary because the noise properties are unique for each CCD. A light curve was created by summing the flux inside the aperture for each of the 15 stacked images. 
To check that the variability seen in the light curve is not due to the choice of aperture, we created light curves using a variety of aperture masks. We used fixed circular apertures, increasing in radius from 1 to 3 pixels, centered on the target position. The resulting light curves were consistent with each other within the uncertainties and they were consistent with the light curve extracted using the SNR-defined apertures. Therefore, we opted for the latter to minimize background noise.

We also performed PRF photometry using the same set of frames that were used to create each stacked image (see Section~\ref{subsec:stacking}). Then we fit the PRF to the flux data as 

\begin{equation}
    y_{[t]} = w_{[t]} \cdot p_{[t]}\ ,
\end{equation}

where $y$ are all the pixel flux values in the set of frames $[t]$, $p$ are the corresponding values of the PRF model at each target's corrected position for the collection of frames, and $w$ are the weights for which we solve using linear modeling. 
These weights are the PRF photometry values at the mid-time of each set of frames.

The results of both the aperture photometry and PRF photometry are presented in Table~\ref{tab:lc} and described in Section~\ref{subsec:res_phot}. The photometric fluxes were converted to TESS magnitudes using a zero-point of $20.44 \pm 0.05$, as provided in the TESS Instrument Handbook\footnote{\url{https://archive.stsci.edu/missions/tess/doc/TESS_Instrument_Handbook_v0.1.pdf}}.

\section{Results}\label{sec:results}

\subsection{Astrometric Precision}\label{subsec:res_astro}

\begin{figure}[h!]
    \centering
    \includegraphics[width=.99\linewidth]{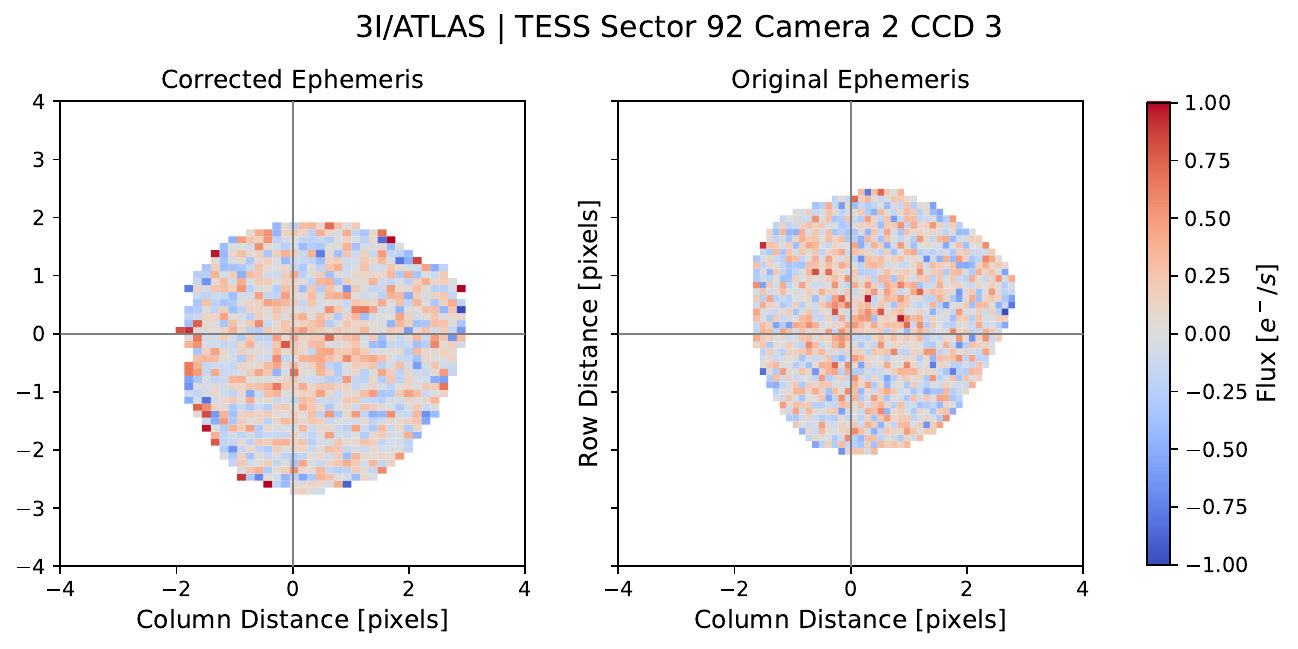}
    \includegraphics[width=.99\linewidth]{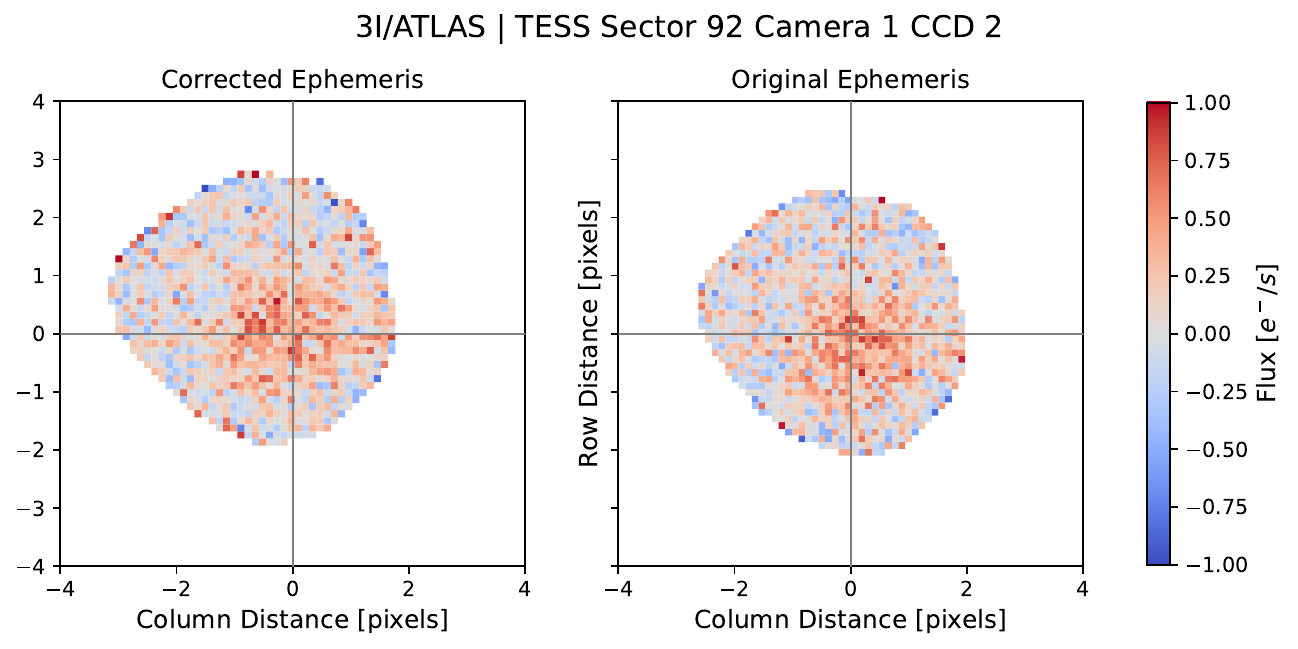}
    \caption{Normalized flux for every pixel in the image cutout with respect to the expected position of \cname. Data correspond to observations from TESS sector 92 camera 2 CCD 3 (top row) and camera 1 CCD 2 (bottom row). The left panels show the original ephemeris from JPL Horizons (accessed on \jplephemdate), while the right panels show the corrected ephemeris as described in Section~\ref{subsec:ephem_corr}. A stronger signal is represented by redder points.}
    \label{fig:offset}
\end{figure}

In Section \ref{subsec:ephem_corr}, we calculated the offset between the observed and expected position of \cname on the TESS detectors, accounting for inaccuracies in both the ephemeris and the TESS WCS. To help visualize our results, Figure~\ref{fig:offset} shows the distribution of pixel flux values as a function of their distance to the target's position, before and after applying our correction model.
The distribution of pixels with significant signal (redder points) are slightly better centered on the origin after applying our model. This is more evident in camera 1 CCD 2 because the target is brighter.
To assess the performance of this model, we computed the flux-weighted average of the distribution in both the column and row axes.
These average values represent the aggregated offset between the expected and true position of the target in pixel space. 
Before our position correction, the weighted average offset for camera 2 CCD 3 is 0.55\,pixels in the column-axis and 0.09 in the row-axis and, after the correction, it is 0.09 in the column-axis and -0.19 in the row-axis. This corresponds to a total offset of 0.56\,pixels before the correction and 0.21\,pixels after the correction.
For camera 1 CCD 2, the values are 0.14 (column-axis) and -0.11 (row-axis) before the correction (total offset 0.17\,pixels) and 0.07 (column-axis) and 0.07 (row-axis) after (total offset 0.10\,pixels).
The results from both CCDs show a significant improvement in the target's position after applying our correction model. In Table~\ref{tab:lc} we report the measured position from our model at the mid-time of each image stack. We converted the pixel coordinates to RA and Dec using the TESS WCS from \texttt{tesswcs} for each camera and CCD. By comparing the measured sky positions to the expected ephemeris from JPL Horizons, we find an average offset of $7.8\arcsec$ in RA and $3.9\arcsec$ in Dec for camera 2 CCD 3, and $2.6\arcsec$ in RA and $3.9\arcsec$ in Dec for camera 1 CCD 2.

We investigated the cause of the offset between the observed position of \cname and the expected position given the JPL Horizons ephemeris. Using a sample of bright ($V_{\text{mag}} < 18$) asteroids with well-constrained ephemerides, we estimated the uncertainty in astrometry originating from the TESS WCS solution. We used \texttt{tess-asteroids} to create TPFs for a sample of 150 asteroids observed on camera 1 CCD 2 (with an average of 642 frames per target) and 34 asteroids observed on camera 2 CCD 3 (with an average of 656 frames per target) during the TESS primary mission. For each frame, we converted the target's predicted position from RA, Dec into row, column using the WCS solution provided by \texttt{tesswcs} and we measured the offset from the flux-weighted centroid inside the target aperture. For camera 2 CCD 3, we found an average offset of $5.0\arcsec$ in RA and $1.9\arcsec$ in Dec. For camera 1 CCD 2, we found an average offset of $2.0\arcsec$ in RA and $0.9\arcsec$ in Dec. These values represent a target's typical offset from its expected position due to inaccuracies in the TESS WCS.

Our correction model marginally improved the positions of \cname in camera 2 CCD 3 and the offsets are slightly larger than expected from inaccuracies in the TESS WCS. This is likely due to the weakly detectable signal in the early observations.
In camera 1 CCD 2, the positions are significantly improved by the correction model. The offset in RA is similar to that expected from the TESS WCS inaccuracy, but the offset in Dec is about four times larger than expected.
Our results indicate that the offset between the expected and measured position is dominated by the uncertainty in the TESS WCS solution rather than uncertainty in the ephemeris of \cname.
Due to the large pixel scale of TESS ($21 \arcsec / \text{pix}$) and the necessity to stack images with hundreds of frames, we are unable to obtain accurate and precise astrometry for \cname and improve its orbit solution.

\subsection{Light Curve}\label{subsec:res_phot}

\begin{figure*}[ht!]
    \centering
    \includegraphics[width=0.48\linewidth]{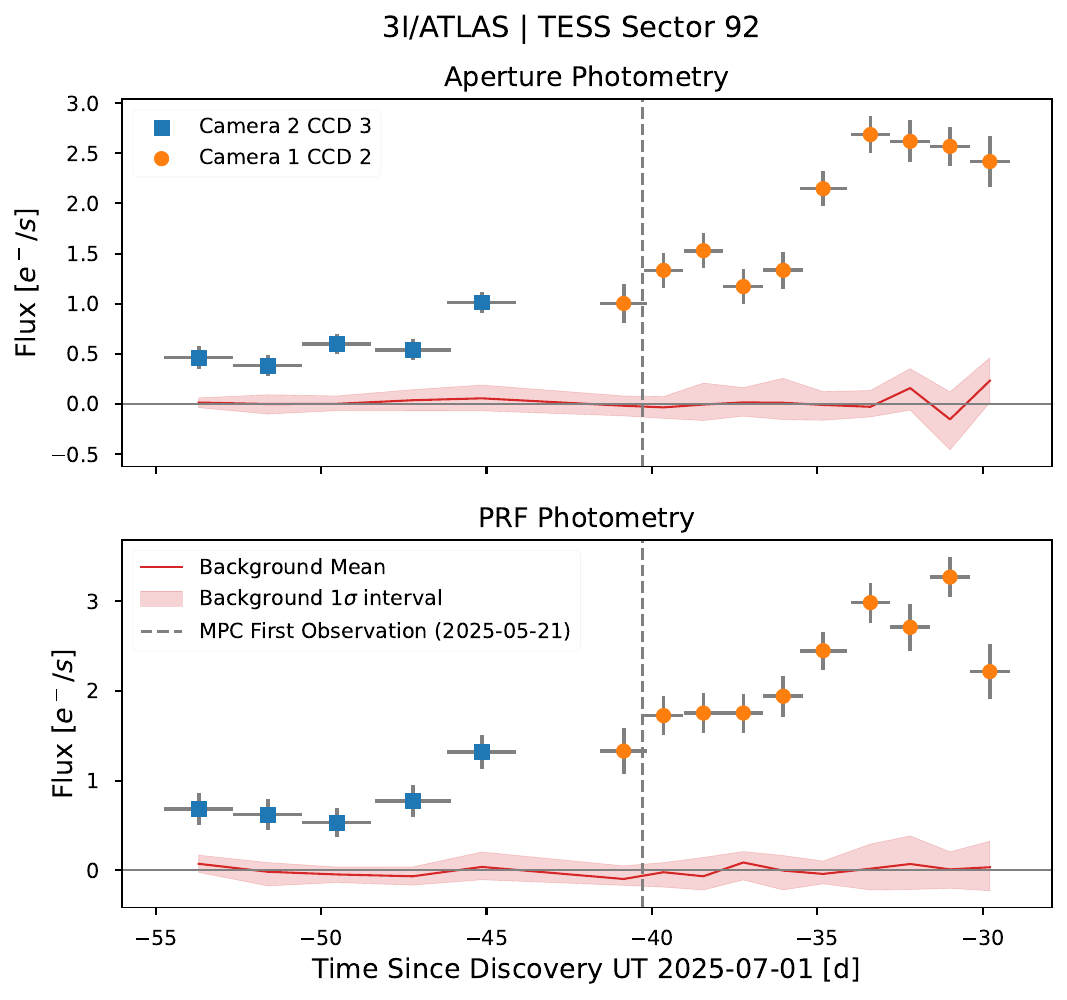}
    \includegraphics[width=0.48\linewidth]{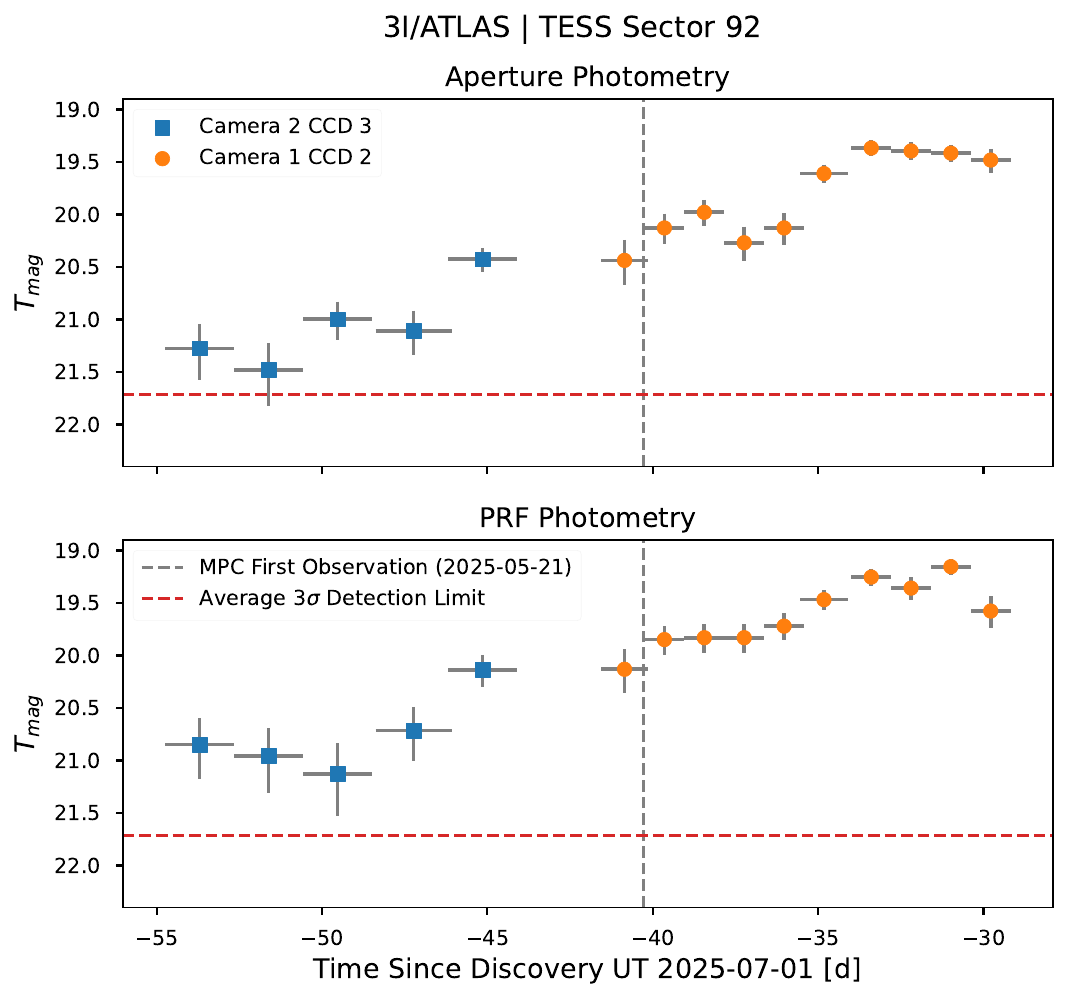}
    \caption{\cname light curves using aperture (top) and PRF (bottom) photometry in flux (left) and magnitude (right), measured from the stacked images. Data from camera 2 CCD 3 is shown in blue squares, while orange circles represent data from camera 1 CCD 2. Error bars in the time axis represent the window used for image stacking. The solid red line and shaded region represent the mean background flux and $1\sigma$ intervals. The horizontal dashed red line represents the $3\sigma$ average detection limit across all stacked images. The x-axis is defined with respect to the time of discovery \discovery.}
    \label{fig:tess_lc}
\end{figure*}

The results of the aperture and PRF photometry are shown in Figure~\ref{fig:tess_lc} and the data is included in Table~\ref{tab:lc}.
Both the aperture and PRF light curves show a similar trend of increasing brightness. The larger scatter and uncertainties in the aperture photometry for camera 2 CCD 3 data is due to residuals from the background subtraction that contribute to the noise, whereas the PRF photometry is less susceptible to this effect.
The PRF light curve of \cname exhibits an increase in brightness from $T_{\text{mag}} \sim 20.9$ in its first observation, 55\,days before discovery (\discovery), to $T_{\text{mag}} \sim 19.6$, 30 days prior.
During this observing window of nearly 26 days, the heliocentric distance of \cname decreased by almost 0.9\,AU, traveling from 6.35 to 5.46\,AU.
Variability analysis during the first part of the observations (camera 2 CCD 3) is not possible due to low brightness, photometric noise and inconsistencies between PRF and aperture photometry.
During the second part of the observations (camera 1 CCD 2), when the signal is stronger, both light curves show a $\sim 0.4$ magnitude increase over 3\,days around day $-35$ which could be due to activity.
Due to the small number of data points in our light curves, a more significant variability analysis is beyond the scope of this work.

We compared our aperture photometry to other light curves to check for consistency. Firstly, we extracted light curves using circular apertures of varying size as described in Section \ref{subsec:phot} (we used the same aperture for both CCDs in these cases). Secondly, we computed a light curve using the camera-specific apertures and image stacks built without our position correction. The aperture photometry light curve presented in this section was consistent with these alternative light curves within the photometric uncertainties. There was a $5\%$ increase in flux and $10\%$ increase in SNR, on average, caused by including the position correction.

\begin{figure*}[ht!]
    \centering
    \includegraphics[width=1\linewidth]{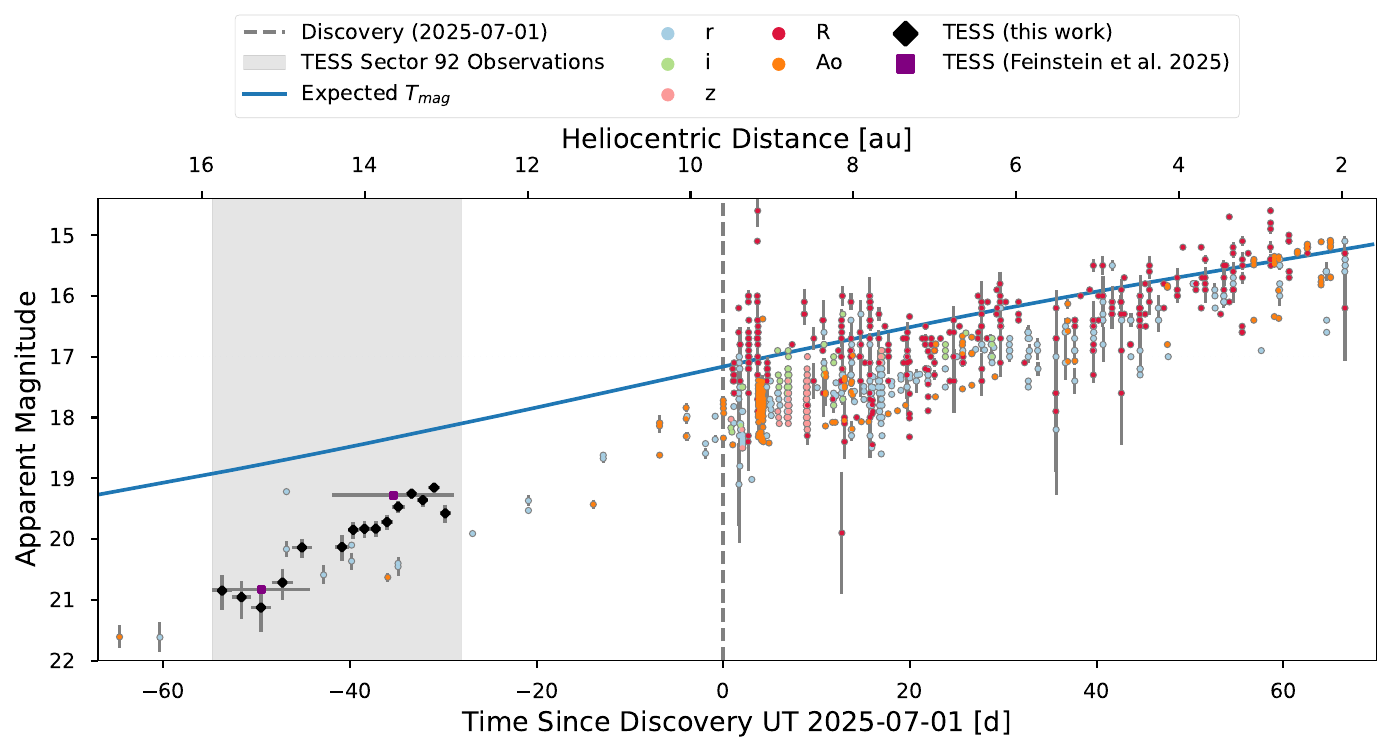}
    \caption{Light curve of \cname from observatories with bandpasses similar to TESS, including the PRF photometry from this work (black diamonds). The TESS measurements from \citet{2025arXiv250721967F} are shown for comparison (purple squares). The top axis shows the heliocentric distance covered by \cname during the observations. The gray shaded region represents the TESS sector 92 observing window. The vertical dashed line represents the date \cname was discovered (\discovery). The blue line shows the expected $T_{\text{mag}}$ as reported by JPL Horizons \citep[converted from $V_{\text{mag}}$ using][]{2021PSJ.....2..236F}. The different colored markers represent photometric bandpasses similar to TESS ($600-1000$\,nm): \textit{r}, \textit{i}, and \textit{z} are equivalent to SDSS \textit{riz} bands; \textit{R} represents the equivalent Johnson-Cousins \textit{R} band (centered on 658\,nm); and \textit{Ao} is ATLAS orange bandpass ($560-820$\,nm). Data was collected from the MPC Explorer service on \mpclcdate \citep{2025MNRAS.542L.139B,2025arXiv250713409C,2025arXiv250800808S,2025MPEC....N..102C,2025MPEC....O...20P,2025arXiv250908792Y,2025arXiv250905562T}.
    }
    \label{fig:mpc_lc}
\end{figure*}

Figure~\ref{fig:mpc_lc} shows photometric measurements of \cname in bandpasses similar to TESS ($600-1000$\,nm) reported to the Minor Planet Center (MPC) as of \mpclcdate. 
The light curve includes our TESS photometry, that from the Vera C. Rubin Observatory \citep{2025arXiv250713409C}, The Zwicky Transient Facility \citep[ZTF, ][]{2025arXiv250908792Y}, and ATLAS \citep{2025arXiv250905562T}.
This highlights the significant contribution of the TESS observations to the sparsely populated pre-discovery region of the light curve.
The TESS photometry matches the increase in brightness of \cname as seen by other observatories.
The measured $T_{\text{mag}}$ is on average 1.5 magnitudes fainter than the expected values reported by JPL Horizons, after converting to the TESS bandpass \citep{2021PSJ.....2..236F}. This is likely due to the sparse amount of pre-discovery data, meaning the model used to extrapolate brightness was poorly constrained in this region.

\begin{figure}[t!]
    \centering
    \includegraphics[width=1\linewidth]{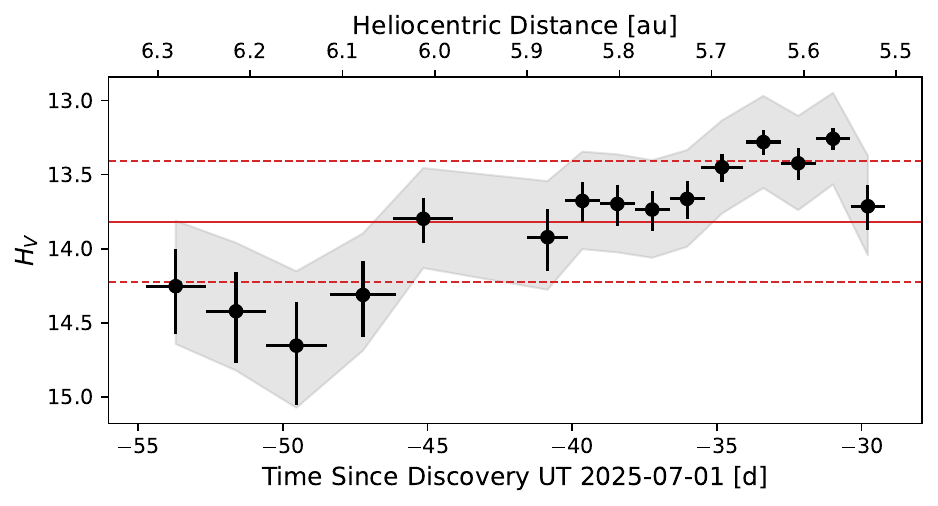}
    \caption{Light curve of \cname in terms of the absolute visual magnitude. The gray shaded region represents the $\pm 0.3$ magnitude uncertainty in the $T_{\text{mag}}$ to $V_{\text{mag}}$ conversion, as reported in \citet{2021PSJ.....2..236F}. The mean absolute magnitude (13.8; solid red horizontal line) and $1\sigma$ bounds ($\pm 0.4$; dashed red horizontal lines) are included for reference. The top axis shows the heliocentric distance covered by \cname during the observations.}
    \label{fig:lc_hv}
\end{figure}

Figure~\ref{fig:lc_hv} shows the light curve of \cname in terms of the visual absolute magnitude, $H_{V}$, also known as a secular light curve. This data is also included in Table~\ref{tab:lc}. To compute this light curve, we converted the PRF magnitudes in the TESS bandpass to magnitudes in the visual bandpass using $V_{\text{mag}} = T_{\text{mag}} + 0.8$ \citep{2021PSJ.....2..236F} with $\pm 0.3$ magnitudes uncertainty, computed for multiple families of comets. We then computed $H_{V}$ according to

\begin{equation}
    H_{V} = V_{\text{mag}} - 2.5 n \text{log}_{10}(d_{\odot}) - 5\text{log}_{10}(d_{\text{obs}})\,
\end{equation}

where $d_{\odot}$ and $d_{\text{obs}}$ are the heliocentric and observer distances, respectively, and $n$ is a comet activity index. We used activity index $n=1.8$ (corresponding to the scaling factor $k1=4.5$) and the object distances at the mid-time of the stacked frames, both reported by JPL Horizons. We find that, during the TESS observations, \cname has an average $H_{V} = 13.8 \pm 0.4$. The average value is in accordance, within $1\sigma$, with contemporaneous observations performed by ZTF \citep[$H_{V} = 13.4 \pm 0.1$;][]{2025arXiv250908792Y}. The variability in the secular light curve could be caused by faint cometary activity, which is consistent with pre-perihelion activity measured by \citet{2025arXiv250800808S, 2025arXiv250818209C,2025ApJ...990L...2J} and the ZTF observations. However, we note that the variability is within the photometric uncertainties, which are dominated by the uncertainty on the zero point when converting $T_{\text{mag}}$ to $V_{\text{mag}}$.

\begin{deluxetable*}{lrllrrrrcllrrrr}
\digitalasset
\tablecaption{Results of Photometry on Stacked Images \label{tab:lc}}
\tablehead{
\colhead{Date Midpoint} & \colhead{Stack Duration} & \colhead{RA} & \colhead{Dec} & \colhead{Ap. Flux} & \colhead{Ap. $T_{\text{mag}}$} & \colhead{PRF Flux} & \colhead{PRF. $T_{\text{mag}}$} & \colhead{SNR} & \colhead{Exp. RA} & \colhead{Exp. Dec} & \colhead{$\Delta$ RA} & \colhead{$\Delta$ Dec} & \colhead{Exp. $T_{\text{mag}}$} & \colhead{$H_{V}$}\\
\colhead{(UT)} & \colhead{(days)} & \colhead{(hhmmss)} & \colhead{(ddmmss)} & \colhead{($e^{-}/s$)}  & \colhead{(mag)} & \colhead{($e^{-}/s$)}  & \colhead{(mag)} & \colhead{(Ap/PRF)} & \colhead{(hhmmss)} & \colhead{(ddmmss)} & \colhead{($\arcsec$)} & \colhead{($\arcsec$)} & \colhead{(mag)} & \colhead{(mag)}
}
\startdata
2025-05-08 12:25:25 & 2.1 & $19^{\mathrm{h}}12^{\mathrm{m}}35^{\mathrm{s}}$ & $-18^\circ42{}^\prime22{}^{\prime\prime}$ & $0.463 \pm 0.111$ & $21.28_{-0.30}^{+0.23}$ & $0.686 \pm 0.177$ & $20.85_{-0.32}^{+0.25}$ & $ 3.2 /  3.6$ & $19^{\mathrm{h}}12^{\mathrm{m}}35^{\mathrm{s}}$ & $-18^\circ42{}^\prime18{}^{\prime\prime}$ & $-10.2\mathrm{{}^{\prime\prime}}$ &   $4.0\mathrm{{}^{\prime\prime}}$ & 18.90 & $14.25_{-0.44}^{+0.39}$\\
2025-05-10 14:29:03 & 2.1 & $19^{\mathrm{h}}11^{\mathrm{m}}27^{\mathrm{s}}$ & $-18^\circ41{}^\prime22{}^{\prime\prime}$ & $0.383 \pm 0.103$ & $21.48_{-0.34}^{+0.26}$ & $0.622 \pm 0.172$ & $20.96_{-0.35}^{+0.27}$ & $ 2.6 /  3.2$ & $19^{\mathrm{h}}11^{\mathrm{m}}26^{\mathrm{s}}$ & $-18^\circ41{}^\prime29{}^{\prime\prime}$ & $-12.3\mathrm{{}^{\prime\prime}}$ &  $-7.2\mathrm{{}^{\prime\prime}}$ & 18.83 & $14.42_{-0.46}^{+0.40}$\\
2025-05-12 16:32:40 & 2.1 & $19^{\mathrm{h}}10^{\mathrm{m}}12^{\mathrm{s}}$ & $-18^\circ40{}^\prime49{}^{\prime\prime}$ & $0.598 \pm 0.100$ & $21.00_{-0.20}^{+0.17}$ & $0.531 \pm 0.165$ & $21.13_{-0.40}^{+0.29}$ & $ 4.1 /  2.8$ & $19^{\mathrm{h}}10^{\mathrm{m}}11^{\mathrm{s}}$ & $-18^\circ40{}^\prime55{}^{\prime\prime}$ & $-13.1\mathrm{{}^{\prime\prime}}$ &  $-6.1\mathrm{{}^{\prime\prime}}$ & 18.77 & $14.65_{-0.50}^{+0.42}$\\
2025-05-14 23:46:18 & 2.1 & $19^{\mathrm{h}}08^{\mathrm{m}}43^{\mathrm{s}}$ & $-18^\circ40{}^\prime42{}^{\prime\prime}$ & $0.541 \pm 0.104$ & $21.11_{-0.23}^{+0.19}$ & $0.776 \pm 0.180$ & $20.72_{-0.29}^{+0.23}$ & $ 3.7 /  4.1$ & $19^{\mathrm{h}}08^{\mathrm{m}}42^{\mathrm{s}}$ & $-18^\circ40{}^\prime44{}^{\prime\prime}$ & $-11.7\mathrm{{}^{\prime\prime}}$ &  $-1.7\mathrm{{}^{\prime\prime}}$ & 18.70 & $14.31_{-0.42}^{+0.38}$\\
2025-05-17 01:49:54 & 2.1 & $19^{\mathrm{h}}07^{\mathrm{m}}17^{\mathrm{s}}$ & $-18^\circ40{}^\prime52{}^{\prime\prime}$ & $1.012 \pm 0.107$ & $20.43_{-0.12}^{+0.11}$ & $1.320 \pm 0.187$ & $20.14_{-0.17}^{+0.14}$ & $ 7.0 /  6.9$ & $19^{\mathrm{h}}07^{\mathrm{m}}16^{\mathrm{s}}$ & $-18^\circ40{}^\prime58{}^{\prime\prime}$ &  $-9.6\mathrm{{}^{\prime\prime}}$ &  $-6.6\mathrm{{}^{\prime\prime}}$ & 18.64 & $13.79_{-0.34}^{+0.33}$\\
\hline
2025-05-21 08:47:03 & 1.2 & $19^{\mathrm{h}}04^{\mathrm{m}}11^{\mathrm{s}}$ & $-18^\circ42{}^\prime05{}^{\prime\prime}$ & $1.003 \pm 0.198$ & $20.44_{-0.24}^{+0.20}$ & $1.329 \pm 0.254$ & $20.13_{-0.23}^{+0.19}$ & $ 6.9 /  6.9$ & $19^{\mathrm{h}}04^{\mathrm{m}}11^{\mathrm{s}}$ & $-18^\circ42{}^\prime18{}^{\prime\prime}$ &  $-6.2\mathrm{{}^{\prime\prime}}$ & $-12.9\mathrm{{}^{\prime\prime}}$ & 18.51 & $13.92_{-0.38}^{+0.36}$\\
2025-05-22 13:43:52 & 1.2 & $19^{\mathrm{h}}03^{\mathrm{m}}13^{\mathrm{s}}$ & $-18^\circ41{}^\prime54{}^{\prime\prime}$ & $1.333 \pm 0.178$ & $20.13_{-0.16}^{+0.14}$ & $1.724 \pm 0.213$ & $19.85_{-0.14}^{+0.13}$ & $ 9.2 /  9.0$ & $19^{\mathrm{h}}03^{\mathrm{m}}13^{\mathrm{s}}$ & $-18^\circ42{}^\prime06{}^{\prime\prime}$ &  $-2.9\mathrm{{}^{\prime\prime}}$ & $-12.4\mathrm{{}^{\prime\prime}}$ & 18.47 & $13.67_{-0.33}^{+0.33}$\\
2025-05-23 18:40:41 & 1.2 & $19^{\mathrm{h}}02^{\mathrm{m}}11^{\mathrm{s}}$ & $-18^\circ41{}^\prime38{}^{\prime\prime}$ & $1.529 \pm 0.177$ & $19.98_{-0.13}^{+0.12}$ & $1.753 \pm 0.224$ & $19.83_{-0.15}^{+0.13}$ & $10.5 /  9.1$ & $19^{\mathrm{h}}02^{\mathrm{m}}11^{\mathrm{s}}$ & $-18^\circ41{}^\prime50{}^{\prime\prime}$ &  $-1.6\mathrm{{}^{\prime\prime}}$ & $-11.2\mathrm{{}^{\prime\prime}}$ & 18.43 & $13.69_{-0.33}^{+0.33}$\\
2025-05-24 23:37:30 & 1.2 & $19^{\mathrm{h}}01^{\mathrm{m}}07^{\mathrm{s}}$ & $-18^\circ41{}^\prime28{}^{\prime\prime}$ & $1.171 \pm 0.177$ & $20.27_{-0.18}^{+0.15}$ & $1.753 \pm 0.217$ & $19.83_{-0.14}^{+0.13}$ & $ 8.0 /  9.2$ & $19^{\mathrm{h}}01^{\mathrm{m}}07^{\mathrm{s}}$ & $-18^\circ41{}^\prime37{}^{\prime\prime}$ &  $-2.0\mathrm{{}^{\prime\prime}}$ &  $-9.6\mathrm{{}^{\prime\prime}}$ & 18.40 & $13.73_{-0.33}^{+0.33}$\\
2025-05-26 04:34:18 & 1.2 & $19^{\mathrm{h}}00^{\mathrm{m}}00^{\mathrm{s}}$ & $-18^\circ41{}^\prime24{}^{\prime\prime}$ & $1.333 \pm 0.182$ & $20.13_{-0.16}^{+0.14}$ & $1.940 \pm 0.228$ & $19.72_{-0.14}^{+0.12}$ & $ 9.2 / 10.1$ & $19^{\mathrm{h}}00^{\mathrm{m}}00^{\mathrm{s}}$ & $-18^\circ41{}^\prime31{}^{\prime\prime}$ &  $-3.3\mathrm{{}^{\prime\prime}}$ &  $-7.8\mathrm{{}^{\prime\prime}}$ & 18.36 & $13.66_{-0.33}^{+0.32}$\\
2025-05-27 09:31:06 & 1.2 & $18^{\mathrm{h}}58^{\mathrm{m}}51^{\mathrm{s}}$ & $-18^\circ41{}^\prime26{}^{\prime\prime}$ & $2.147 \pm 0.174$ & $19.61_{-0.09}^{+0.08}$ & $2.448 \pm 0.211$ & $19.47_{-0.10}^{+0.09}$ & $14.8 / 12.8$ & $18^{\mathrm{h}}58^{\mathrm{m}}51^{\mathrm{s}}$ & $-18^\circ41{}^\prime33{}^{\prime\prime}$ &  $-4.9\mathrm{{}^{\prime\prime}}$ &  $-6.2\mathrm{{}^{\prime\prime}}$ & 18.32 & $13.44_{-0.32}^{+0.31}$\\
2025-05-28 19:49:35 & 1.2 & $18^{\mathrm{h}}57^{\mathrm{m}}26^{\mathrm{s}}$ & $-18^\circ41{}^\prime39{}^{\prime\prime}$ & $2.686 \pm 0.188$ & $19.37_{-0.08}^{+0.07}$ & $2.983 \pm 0.223$ & $19.25_{-0.08}^{+0.08}$ & $18.5 / 15.6$ & $18^{\mathrm{h}}57^{\mathrm{m}}26^{\mathrm{s}}$ & $-18^\circ41{}^\prime44{}^{\prime\prime}$ &  $-6.4\mathrm{{}^{\prime\prime}}$ &  $-4.8\mathrm{{}^{\prime\prime}}$ & 18.27 & $13.28_{-0.31}^{+0.31}$\\
2025-05-30 00:43:02 & 1.2 & $18^{\mathrm{h}}56^{\mathrm{m}}13^{\mathrm{s}}$ & $-18^\circ41{}^\prime58{}^{\prime\prime}$ & $2.618 \pm 0.209$ & $19.39_{-0.09}^{+0.08}$ & $2.708 \pm 0.265$ & $19.36_{-0.11}^{+0.10}$ & $18.0 / 14.1$ & $18^{\mathrm{h}}56^{\mathrm{m}}13^{\mathrm{s}}$ & $-18^\circ42{}^\prime02{}^{\prime\prime}$ &  $-6.5\mathrm{{}^{\prime\prime}}$ &  $-4.4\mathrm{{}^{\prime\prime}}$ & 18.23 & $13.42_{-0.32}^{+0.32}$\\
2025-05-31 05:36:29 & 1.2 & $18^{\mathrm{h}}54^{\mathrm{m}}58^{\mathrm{s}}$ & $-18^\circ42{}^\prime25{}^{\prime\prime}$ & $2.569 \pm 0.195$ & $19.42_{-0.09}^{+0.08}$ & $3.268 \pm 0.227$ & $19.15_{-0.08}^{+0.07}$ & $17.7 / 17.1$ & $18^{\mathrm{h}}54^{\mathrm{m}}58^{\mathrm{s}}$ & $-18^\circ42{}^\prime30{}^{\prime\prime}$ &  $-4.8\mathrm{{}^{\prime\prime}}$ &  $-5.2\mathrm{{}^{\prime\prime}}$ & 18.20 & $13.25_{-0.31}^{+0.31}$\\
2025-06-01 10:29:55 & 1.2 & $18^{\mathrm{h}}53^{\mathrm{m}}41^{\mathrm{s}}$ & $-18^\circ43{}^\prime00{}^{\prime\prime}$ & $2.417 \pm 0.252$ & $19.48_{-0.12}^{+0.11}$ & $2.215 \pm 0.309$ & $19.58_{-0.16}^{+0.14}$ & $16.6 / 11.6$ & $18^{\mathrm{h}}53^{\mathrm{m}}41^{\mathrm{s}}$ & $-18^\circ43{}^\prime08{}^{\prime\prime}$ &  $-0.3\mathrm{{}^{\prime\prime}}$ &  $-7.6\mathrm{{}^{\prime\prime}}$ & 18.16 & $13.71_{-0.34}^{+0.33}$\\
\enddata
\tablecomments{
The first 5 rows correspond to observations in sector 92 camera 2 CCD 3 and the remaining 10 rows are from sector 92 camera 1 CCD 2.
\textit{Date Midpoint} is computed accounting for all stacked frames, time is given in UT. 
\textit{Stack Duration} corresponds to the duration of the time window used for image stacking. \textit{RA} and \textit{Dec} are the measured object positions from our model, as described in Section~\ref{subsec:ephem_corr}. 
\textit{SNR} is computed as the aggregated flux within the aperture or the PRF flux of the target over the the standard deviation of the background. The \textit{Exp. RA} and \textit{Exp. Dec} are the expected values from the JPL Horizons system. The \textit{$\Delta$ RA} and \textit{$\Delta$ Dec} correspond to the offset between the measured and expected coordinates. 
\textit{Exp.} $T_{\text{mag}}$ is the expected TESS magnitude, calculated using the comet apparent visual total magnitude from JPL Horizons and the conversion from \citet{2021PSJ.....2..236F}. $H_{V}$ is the absolute visual magnitude as described in Section~\ref{subsec:res_phot}.
}
\end{deluxetable*}


\section{Discussion and Conclusions} \label{sec:disc_conc}

We report TESS observations of \cname starting 54.7 days before its discovery (\discovery).
TESS observed \cname for 26 days during sector 92, with a cadence of 200\,seconds. The target was observed by two different CCDs during this time.
Due to the object's low brightness and the fast cadence observations, single frame detections are not possible.
Performing photometry on individual frames is heavily influenced by the quality of the background model and dominated by background noise.

We use linear modeling to shift the ephemeris of \cname to match the TESS observations.
With this correction model we were able to improve the averaged centroid of the detection, particularly in camera 1 CCD 2 where the target's signal is stronger. 
This resulted in average coordinate offsets between the ephemeris prediction and measured positions on the order of the TESS astrometric precision computed from a sample of well-characterized asteroids.
However, the astrometry computed for \cname from these TESS observations is not precise enough to improve the orbit solution.

We perform tracked image stacking, that includes our position corrections, to improve the target's SNR.
We computed aperture photometry, with an aperture mask defined for each camera/CCD, on the resulting stacked images. Additionally, we computed PRF photometry on the same data.
The result is 15 photometric measurements with $\text{SNR}>3$ during the observation window. 
The TESS PRF photometry shows a steady increase in the brightness of \cname, from $T_{\text{mag}} = 20.9 \pm 0.3$ to $T_{\text{mag}} = 19.6 \pm 0.2$, matching the trend seen by subsequent observations with other observatories.
The secular light curve shows evidence of early, weak cometary activity in the TESS observations. This is consistent with recent findings of dust outflows as early as $d_{\odot} \sim 9$\,AU.
The astrometry and photometry measured in this work were reported to the MPC and have been included in the database.

The pre-discovery TESS observations of \cname add crucial data points to an otherwise sparsely populated region of its light curve.
We have verified the target's ephemeris and confirmed the trend in brightness seen by other facilities. 
These observations of \cname as it approaches the inner Solar System will complement future observations as the comet departs in 2026. A complete chronological characterization of \cname will provide further constraints on its dynamical properties, brightness evolution, chemical composition, and activity-driving mechanisms.

Current algorithms to detect moving objects in large sky surveys are predominantly based on assuming the object's direction of movement and speed in order to reduce computational costs.
Due to the interstellar nature of this object, its unique orbit, and low brightness, early discovery in TESS data is significantly harder for image-based detection methods such as shift and stack. 
For example, the state-of-the-art asteroid detection method implemented for TESS data in \citet{2024AJ....167..113N} would have detected \cname in the early days of sector 92 with a probability lower than 40\%.
This urges the necessity to explore new detection algorithms that are better suited to work with the current volume of data produced by large sky surveys such as TESS, the Vera C. Rubin Observatory, and the future Nancy Grace Roman Space Telescope.
Early detection of interesting asteroids, including near-Earth asteroids or interstellar objects, is also limited by TESS operations, such as the data downlink schedule and data processing times. Although \cname was observed in the first days of sector 92 (starting \tessfirst), its detection would only have been hypothetically possible after \mbox{2025-05-22} when the first batch of calibrated FFIs were released.

Finally, with its current orbit, \cname will not be observed again by TESS during year 8 (\mbox{2025-09-15}~--~\mbox{2026-09-07}).
Even though TESS is not designed to study small bodies in our Solar System, it can still provide valuable information as demonstrated by this work. This will become more relevant, as the Vera C. Rubin Observatory is expected to discover more interstellar objects in the upcoming years. TESS will continue to make valuable contributions to the characterization of small bodies and interstellar objects, as well as to planetary defense efforts, thanks to its short cadence time-domain observations and large survey area which are complementary to other ground- and space-based observatories. 

We note that an independent analysis of the TESS observations of \cname was submitted contemporaneously with this work \citep{2025arXiv250721967F}, however a detailed comparison of the two studies is beyond the scope of this paper.


\begin{acknowledgments}

This paper includes data collected by the TESS mission. Funding for the TESS mission is provided by the NASA's Science Mission Directorate.
This paper includes data obtained from the MAST data archive at the Space Telescope Science Institute (STScI). STScI is operated by the Association of Universities for Research in Astronomy, Inc., under NASA contract NAS 5–26555.
Funding for this work for JMP, AT and CH is provided by NASA grant 80NSSC20M0192. 
The material is based upon work supported by NASA under award number 80GSFC24M0006.
This research made use of \texttt{Lightkurve}, a Python package for Kepler and TESS data analysis \citep{2018ascl.soft12013L}.

\end{acknowledgments}

\facilities{TESS}

\software{
    \texttt{astropy} \citep{astropy:2013, astropy:2018, astropy:2022},
    \texttt{lkprf} (\url{https://github.com/lightkurve/lkprf}),
    \texttt{lkspacecraft} \citep{Hedges_2025},
    \texttt{matplotlib} \citep{matplotlib},
    \texttt{numpy} \citep{harris2020array},
    \texttt{tess-asteroids} \citep{tuson_2025_16332750},
    \texttt{tess-ephem} (\url{https://github.com/SSDataLab/tess-ephem}),
    \texttt{tesscube} (\url{https://github.com/tessgi/tesscube}),
    \texttt{tesswcs} (\url{https://github.com/tessgi/tesswcs}).
    }

\bibliography{bibliography}{}

\begin{thebibliography}{}
\expandafter\ifx\csname natexlab\endcsname\relax\def\natexlab#1{#1}\fi
\providecommand{\url}[1]{\href{#1}{#1}}
\providecommand{\dodoi}[1]{doi:~\href{http://doi.org/#1}{\nolinkurl{#1}}}
\providecommand{\doeprint}[1]{\href{http://ascl.net/#1}{\nolinkurl{http://ascl.net/#1}}}
\providecommand{\doarXiv}[1]{\href{https://arxiv.org/abs/#1}{\nolinkurl{https://arxiv.org/abs/#1}}}

\bibitem[{ {Astropy Collaboration} {et~al.}(2013){Astropy Collaboration}, {Robitaille}, {Tollerud}, {Greenfield}, {Droettboom}, {Bray}, {Aldcroft}, {Davis}, {Ginsburg}, {Price-Whelan}, {Kerzendorf}, {Conley}, {Crighton}, {Barbary}, {Muna}, {Ferguson}, {Grollier}, {Parikh}, {Nair}, {Unther}, {Deil}, {Woillez}, {Conseil}, {Kramer}, {Turner}, {Singer}, {Fox}, {Weaver}, {Zabalza}, {Edwards}, {Azalee Bostroem}, {Burke}, {Casey}, {Crawford}, {Dencheva}, {Ely}, {Jenness}, {Labrie}, {Lim}, {Pierfederici}, {Pontzen}, {Ptak}, {Refsdal}, {Servillat}, \& {Streicher}}]{astropy:2013}
{Astropy Collaboration}, {Robitaille}, T.~P., {Tollerud}, E.~J., {et~al.} 2013, \bibinfo{title}{{Astropy: A community Python package for astronomy},} \aap, 558, A33, \dodoi{10.1051/0004-6361/201322068}

\bibitem[{ {Astropy Collaboration} {et~al.}(2018){Astropy Collaboration}, {Price-Whelan}, {Sip{\H{o}}cz}, {G{\"u}nther}, {Lim}, {Crawford}, {Conseil}, {Shupe}, {Craig}, {Dencheva}, {Ginsburg}, {Vand erPlas}, {Bradley}, {P{\'e}rez-Su{\'a}rez}, {de Val-Borro}, {Aldcroft}, {Cruz}, {Robitaille}, {Tollerud}, {Ardelean}, {Babej}, {Bach}, {Bachetti}, {Bakanov}, {Bamford}, {Barentsen}, {Barmby}, {Baumbach}, {Berry}, {Biscani}, {Boquien}, {Bostroem}, {Bouma}, {Brammer}, {Bray}, {Breytenbach}, {Buddelmeijer}, {Burke}, {Calderone}, {Cano Rodr{\'\i}guez}, {Cara}, {Cardoso}, {Cheedella}, {Copin}, {Corrales}, {Crichton}, {D'Avella}, {Deil}, {Depagne}, {Dietrich}, {Donath}, {Droettboom}, {Earl}, {Erben}, {Fabbro}, {Ferreira}, {Finethy}, {Fox}, {Garrison}, {Gibbons}, {Goldstein}, {Gommers}, {Greco}, {Greenfield}, {Groener}, {Grollier}, {Hagen}, {Hirst}, {Homeier}, {Horton}, {Hosseinzadeh}, {Hu}, {Hunkeler}, {Ivezi{\'c}}, {Jain}, {Jenness}, {Kanarek}, {Kendrew}, {Kern}, {Kerzendorf}, {Khvalko}, {King}, {Kirkby}, {Kulkarni},
  {Kumar}, {Lee}, {Lenz}, {Littlefair}, {Ma}, {Macleod}, {Mastropietro}, {McCully}, {Montagnac}, {Morris}, {Mueller}, {Mumford}, {Muna}, {Murphy}, {Nelson}, {Nguyen}, {Ninan}, {N{\"o}the}, {Ogaz}, {Oh}, {Parejko}, {Parley}, {Pascual}, {Patil}, {Patil}, {Plunkett}, {Prochaska}, {Rastogi}, {Reddy Janga}, {Sabater}, {Sakurikar}, {Seifert}, {Sherbert}, {Sherwood-Taylor}, {Shih}, {Sick}, {Silbiger}, {Singanamalla}, {Singer}, {Sladen}, {Sooley}, {Sornarajah}, {Streicher}, {Teuben}, {Thomas}, {Tremblay}, {Turner}, {Terr{\'o}n}, {van Kerkwijk}, {de la Vega}, {Watkins}, {Weaver}, {Whitmore}, {Woillez}, {Zabalza}, \& {Astropy Contributors}}]{astropy:2018}
{Astropy Collaboration}, {Price-Whelan}, A.~M., {Sip{\H{o}}cz}, B.~M., {et~al.} 2018, \bibinfo{title}{{The Astropy Project: Building an Open-science Project and Status of the v2.0 Core Package},} \aj, 156, 123, \dodoi{10.3847/1538-3881/aabc4f}

\bibitem[{ {Astropy Collaboration} {et~al.}(2022){Astropy Collaboration}, {Price-Whelan}, {Lim}, {Earl}, {Starkman}, {Bradley}, {Shupe}, {Patil}, {Corrales}, {Brasseur}, {N{"o}the}, {Donath}, {Tollerud}, {Morris}, {Ginsburg}, {Vaher}, {Weaver}, {Tocknell}, {Jamieson}, {van Kerkwijk}, {Robitaille}, {Merry}, {Bachetti}, {G{"u}nther}, {Aldcroft}, {Alvarado-Montes}, {Archibald}, {B{'o}di}, {Bapat}, {Barentsen}, {Baz{'a}n}, {Biswas}, {Boquien}, {Burke}, {Cara}, {Cara}, {Conroy}, {Conseil}, {Craig}, {Cross}, {Cruz}, {D'Eugenio}, {Dencheva}, {Devillepoix}, {Dietrich}, {Eigenbrot}, {Erben}, {Ferreira}, {Foreman-Mackey}, {Fox}, {Freij}, {Garg}, {Geda}, {Glattly}, {Gondhalekar}, {Gordon}, {Grant}, {Greenfield}, {Groener}, {Guest}, {Gurovich}, {Handberg}, {Hart}, {Hatfield-Dodds}, {Homeier}, {Hosseinzadeh}, {Jenness}, {Jones}, {Joseph}, {Kalmbach}, {Karamehmetoglu}, {Ka{l}uszy{'n}ski}, {Kelley}, {Kern}, {Kerzendorf}, {Koch}, {Kulumani}, {Lee}, {Ly}, {Ma}, {MacBride}, {Maljaars}, {Muna}, {Murphy}, {Norman}, {O'Steen},
  {Oman}, {Pacifici}, {Pascual}, {Pascual-Granado}, {Patil}, {Perren}, {Pickering}, {Rastogi}, {Roulston}, {Ryan}, {Rykoff}, {Sabater}, {Sakurikar}, {Salgado}, {Sanghi}, {Saunders}, {Savchenko}, {Schwardt}, {Seifert-Eckert}, {Shih}, {Jain}, {Shukla}, {Sick}, {Simpson}, {Singanamalla}, {Singer}, {Singhal}, {Sinha}, {Sip{H{o}}cz}, {Spitler}, {Stansby}, {Streicher}, {{{S}}umak}, {Swinbank}, {Taranu}, {Tewary}, {Tremblay}, {Val-Borro}, {Van Kooten}, {Vasovi{'c}}, {Verma}, {de Miranda Cardoso}, {Williams}, {Wilson}, {Winkel}, {Wood-Vasey}, {Xue}, {Yoachim}, {Zhang}, {Zonca}, \& {Astropy Project Contributors}}]{astropy:2022}
{Astropy Collaboration}, {Price-Whelan}, A.~M., {Lim}, P.~L., {et~al.} 2022, \bibinfo{title}{{The Astropy Project: Sustaining and Growing a Community-oriented Open-source Project and the Latest Major Release (v5.0) of the Core Package},} \apj, 935, 167, \dodoi{10.3847/1538-4357/ac7c74}

\bibitem[{B.~T. {Bolin} {et~al.}(2025){Bolin}, {Belyakov}, {Fremling}, {Graham}, {Abdelaziz}, {Elhosseiny}, {Gray}, {Ingebretsen}, {Jewett}, {Lisse}, {Karpov}, {Kilic}, {Ma{\v{s}}ek}, {Molham}, {Roderick}, {Takey}, {Abron}, {Coughlin}, {Hsieh}, {Noll}, \& {Wong}}]{2025MNRAS.542L.139B}
{Bolin}, B.~T., {Belyakov}, M., {Fremling}, C., {et~al.} 2025, \bibinfo{title}{{Interstellar comet 3I/ATLAS: discovery and physical description},} \mnras, 542, L139, \dodoi{10.1093/mnrasl/slaf078}

\bibitem[{G. {Borisov} {et~al.}(2019){Borisov}, {Durig}, {Sato}, {et~al.}}]{2019CBET.4666}
{Borisov}, G., {Durig}, D.~T., {Sato}, H., {et~al.} 2019, \bibinfo{title}{{COMET C/2019 Q4 (BORISOV)},} Central Bureau Electronic Telegrams, 4666, 1

\bibitem[{C.~O. {Chandler} {et~al.}(2025){Chandler}, {Bernardinelli}, {Juri{\'c}}, {Singh}, {Hsieh}, {Sullivan}, {Jones}, {Kurlander}, {Vavilov}, {Eggl}, {Holman}, {Spoto}, {Schwamb}, {Christensen}, {Beebe}, {Roodman}, {Lim}, {Jenness}, {Bosch}, {Smart}, {Bellm}, {MacBride}, {Rawls}, {Greenstreet}, {Slater}, {Heinze}, {Ivezi{\'c}}, {Blum}, {Connolly}, {Daues}, {Makadia}, {Gower}, {Bryce Kalmbach}, {Monet}, {Bannister}, {Dones}, {Dorsey}, {Fraser}, {Forbes}, {Fuentes}, {Holt}, {Inno}, {Jones}, {Knight}, {Lintott}, {Lister}, {Lupton}, {Mendoza Magbanua}, {Malhotra}, {Mueller}, {Murtagh}, {Pandey}, {Reach}, {Samarasinha}, {Seligman}, {Snodgrass}, {Solontoi}, {Szab{\'o}}, {White}, {Womack}, {Young}, {Allbery}, {Armellin}, {Aubourg}, {Avdellidou}, {Azfar}, {Bauer}, {Bechtol}, {Belyakov}, {Benecchi}, {Bertini}, {Bolin}, {Bose}, {Buchanan}, {Boucaud}, {Boufleur}, {Boutigny}, {Braga-Ribas}, {Calabrese}, {Camargo}, {Caplar}, {Carry}, {Carvajal}, {Choi}, {Cowan}, {Croft}, {{\'C}uk}, {Daruich}, {Daubard}, {Davenport},
  {Daylan}, {Delgado}, {Devillepoix}, {Doherty}, {Donaldson}, {Drass}, {Deppe}, {Dubois-Felsmann}, {Economou}, {Eduardo}, {Farnocchia}, {Frissell}, {Fedorets}, {Fernandes}, {Fulle}, {Gerdes}, {Gibbs}, {Gillan}, {Guy}, {Hammergren}, {Hanushevsky}, {Hernandez}, {Hestroffer}, {Hopkins}, {Granvik}, {Ieva}, {Irving}, {Jannuzi}, {Jimenez}, {Ramos Gomes-J{\'u}nior}, {Juramy}, {Kahn}, {Kannawadi}, {Kang}, {Kryszczy{\'n}ska}, {Kotov}, {Koumjian}, {Krughoff}, {Lage}, {Lange}, {Levine}, {Li}, {Licandro}, {Lin}, {Lust}, {Lyttle}, {Mahabal}, {Mahlke}, {Plazas Malag{\'o}n}, {Salazar Manzano}, {Marc}, {Margoti}, {Mar{\v{c}}eta}, {Menanteau}, {Meyers}, {Mills}, {Morato}, {More}, {Morrison}, {Moulane}, {Mu{\~n}oz-Guti{\'e}rrez}, {M.}, {O'Connor}, {Oldag}, {Oldroyd}, {O'Mullane}, {Opitom}, {Oszkiewicz}, {Page}, {Patterson}, {Payne}, {Peloton}, {Pereira}, {Peterson}, {Polin}, {Pollek}, {Polen}, {Qiu}, {Ragozzine}, {Rajagopal}, {van Reeven}, {Rice}, {Ridgway}, {Rivkin}, {Robinson}, {Ro{\.z}ek}, {Salnikov}, {S{\'a}nchez},
  {Sarid}, {Schambeau}, {Scolnic}, {Schindler}, {Seaman}, {Jacques}, {Shaw}, {Shugart}, {Sick}, {Siraj}, {Sitarz}, {Sobhani}, {Soldahl}, {Stalder}, {Stetzler}, {Swinbank}, {Szigeti}, {Tauraso}, {Thornton}, {Tonietti}, {Trilling}, \& {Trujillo}}]{2025arXiv250713409C}
{Chandler}, C.~O., {Bernardinelli}, P.~H., {Juri{\'c}}, M., {et~al.} 2025, \bibinfo{title}{{NSF-DOE Vera C. Rubin Observatory Observations of Interstellar Comet 3I/ATLAS (C/2025 N1)},} arXiv e-prints, arXiv:2507.13409.
\newblock \doarXiv{2507.13409}

\bibitem[{A. {Chernyshov} {et~al.}(2025){Chernyshov}, {Nazarov}, {Robinson}, {Tonry}, {Weiland}, {Siverd}, {Erasmus}, {Fitzsimmons}, {Denneau}, {James}, {:unkn}, {Ivanov}, {Ivanova}, {Barcov}, {Roshchupko}, {Lysenko}, {Kurbatov}, {Archibasov}, {Ivanov}, {Yakovenko}, {Ozhered}, {Masi}, {Diepvens}, {Iozzi}, {Bernardi}, {Tombelli}, {Lombardo}, {Interrante}, {Mazzanti}, {Birtwhistle}, {Ferreira}, {Gomez}, {Lister}, {Greenstreet}, {Holt}, {Chatelain}, {Merlin}, {Ruocco}, {Theron}, {Bacci}, {Jacques}, {Laborde}, {Dupouy}, {Sato}, {Ursache}, {Jaeger}, {Rhemann}, {Prosperi}, {Elek}, {Grazzini}, {Camarasa}, {Kugel}, {Reina}, {Koch}, {Campas}, {Naves}, {Gerhard}, \& {Hug}}]{2025MPEC....N..102C}
{Chernyshov}, A., {Nazarov}, S., {Robinson}, J., {et~al.} 2025, \bibinfo{title}{{Comet 3I/ATLAS},} Minor Planet Electronic Circulars, 2025-N102, \dodoi{10.48377/MPEC/2025-N102}

\bibitem[{M.~A. {Cordiner} {et~al.}(2025){Cordiner}, {Roth}, {Kelley}, {Bodewits}, {Charnley}, {Drozdovskaya}, {Farnocchia}, {Micheli}, {Milam}, {Opitom}, {Schwamb}, {Thomas}, \& {Bagnulo}}]{2025arXiv250818209C}
{Cordiner}, M.~A., {Roth}, N.~X., {Kelley}, M. S.~P., {et~al.} 2025, \bibinfo{title}{{JWST detection of a carbon dioxide dominated gas coma surrounding interstellar object 3I/ATLAS},} arXiv e-prints, arXiv:2508.18209, \dodoi{10.48550/arXiv.2508.18209}

\bibitem[{R. {de la Fuente Marcos} {et~al.}(2025){de la Fuente Marcos}, {Licandro}, {Alarcon}, {Serra-Ricart}, {de Leon}, {de la Fuente Marcos}, {Lombardi}, {Tejero}, {Cabrera-Lavers}, {Guerra Arencibia}, \& {Ruiz Cejudo}}]{2025arXiv250712922D}
{de la Fuente Marcos}, R., {Licandro}, J., {Alarcon}, M.~R., {et~al.} 2025, \bibinfo{title}{{Assessing interstellar comet 3I/ATLAS with the 10.4 m Gran Telescopio Canarias and the Two-meter Twin Telescope},} arXiv e-prints, arXiv:2507.12922, \dodoi{10.48550/arXiv.2507.12922}

\bibitem[{L. {Denneau} {et~al.}(2025){Denneau}, {Siverd}, {Tonry}, {Weiland}, {Erasmus}, {Fitzsimmons}, {Robinson}, {Deen}, {Collaboration}, {Ye}, {Gilmore}, {Kilmartin}, {Ursache}, {Korlevic}, {Poropat}, {Urbanik}, {Prosperi}, {Jaeger}, {Rhemann}, {Duin}, {Hale}, {Moravec}, {Jacques}, {Neue}, {Alarcon}, {Licandro}, {Nichita}, {Ly}, {Schnabel}, {Bamberger}, {Ruhela}, {Serra-Ricart}, {Alarcon}, {Cortes}, {Parrott}, {Rankin}, {Sato}, {Romanov}, {Hutton}, {Linder}, {Holmes}, {Masek}, {Oca{\~n}a}, {Conversi}, {Kresken}, {Micheli}, {Devogele}, {Santana-Ros}, {Lister}, {Greenstreet}, {Holt}, {Gomez}, {Chatelain}, {Manset}, {Silva}, {Weryk}, {Wainscoat}, {Rocchetto}, {Ferguson}, {Guido}, {Serebryanskiy}, {Reva}, \& {Hudin}}]{2025MPEC....N...12D}
{Denneau}, L., {Siverd}, R., {Tonry}, J., {et~al.} 2025, \bibinfo{title}{{3I/ATLAS = C/2025 N1 (ATLAS)},} Minor Planet Electronic Circulars, 2025-N12, \dodoi{10.48377/MPEC/2025-N12}

\bibitem[{T.~L. {Farnham} {et~al.}(2021){Farnham}, {Kelley}, \& {Bauer}}]{2021PSJ.....2..236F}
{Farnham}, T.~L., {Kelley}, M. S.~P., \& {Bauer}, J.~M. 2021, \bibinfo{title}{{Early Activity in Comet C/2014 UN271 Bernardinelli-Bernstein as Observed by TESS},} \psj, 2, 236, \dodoi{10.3847/PSJ/ac323d}

\bibitem[{T.~L. {Farnham} {et~al.}(2019){Farnham}, {Kelley}, {Knight}, \& {Feaga}}]{2019ApJ...886L..24F}
{Farnham}, T.~L., {Kelley}, M. S.~P., {Knight}, M.~M., \& {Feaga}, L.~M. 2019, \bibinfo{title}{{First Results from TESS Observations of Comet 46P/Wirtanen},} \apjl, 886, L24, \dodoi{10.3847/2041-8213/ab564d}

\bibitem[{A.~D. {Feinstein} {et~al.}(2025){Feinstein}, {Noonan}, \& {Seligman}}]{2025arXiv250721967F}
{Feinstein}, A.~D., {Noonan}, J.~W., \& {Seligman}, D.~Z. 2025, \bibinfo{title}{{Precovery Observations of 3I/ATLAS from TESS Suggests Possible Distant Activity},} arXiv e-prints, arXiv:2507.21967.
\newblock \doarXiv{2507.21967}

\bibitem[{A. {Fitzsimmons} {et~al.}(2023){Fitzsimmons}, {Meech}, {Matr{\`a}}, \& {Pfalzner}}]{2023arXiv230317980F}
{Fitzsimmons}, A., {Meech}, K., {Matr{\`a}}, L., \& {Pfalzner}, S. 2023, \bibinfo{title}{{Interstellar Objects and Exocomets},} arXiv e-prints, arXiv:2303.17980, \dodoi{10.48550/arXiv.2303.17980}

\bibitem[{N. {Halko} {et~al.}(2009){Halko}, {Martinsson}, \& {Tropp}}]{2009arXiv0909.4061H}
{Halko}, N., {Martinsson}, P.-G., \& {Tropp}, J.~A. 2009, \bibinfo{title}{{Finding structure with randomness: Probabilistic algorithms for constructing approximate matrix decompositions},} arXiv e-prints, arXiv:0909.4061, \dodoi{10.48550/arXiv.0909.4061}

\bibitem[{C.~R. Harris {et~al.}(2020)Harris, Millman, van~der Walt, Gommers, Virtanen, Cournapeau, Wieser, Taylor, Berg, Smith, Kern, Picus, Hoyer, van Kerkwijk, Brett, Haldane, del R{\'{i}}o, Wiebe, Peterson, G{\'{e}}rard-Marchant, Sheppard, Reddy, Weckesser, Abbasi, Gohlke, \& Oliphant}]{harris2020array}
Harris, C.~R., Millman, K.~J., van~der Walt, S.~J., {et~al.} 2020, \bibinfo{title}{Array programming with {NumPy},} Nature, 585, 357, \dodoi{10.1038/s41586-020-2649-2}

\bibitem[{C. Hedges {et~al.}(2025)Hedges, Martínez-Palomera, Tuson, Pritchard, Hounsell, Schanche, Kostov, \& Giles}]{Hedges_2025}
Hedges, C., Martínez-Palomera, J., Tuson, A., {et~al.} 2025, \bibinfo{title}{lkspacecraft: A Tool for Obtaining Orbital Properties of the Kepler and TESS Spacecraft,} Research Notes of the AAS, 9, 186, \dodoi{10.3847/2515-5172/adef3a}

\bibitem[{J.~D. Hunter(2007)Hunter}]{matplotlib}
Hunter, J.~D. 2007, \bibinfo{title}{Matplotlib: A 2D graphics environment,} Computing in Science \& Engineering, 9, 90, \dodoi{10.1109/MCSE.2007.55}

\bibitem[{J.~M. {Jenkins} {et~al.}(2016){Jenkins}, {Twicken}, {McCauliff}, {Campbell}, {Sanderfer}, {Lung}, {Mansouri-Samani}, {Girouard}, {Tenenbaum}, {Klaus}, {Smith}, {Caldwell}, {Chacon}, {Henze}, {Heiges}, {Latham}, {Morgan}, {Swade}, {Rinehart}, \& {Vanderspek}}]{spoc}
{Jenkins}, J.~M., {Twicken}, J.~D., {McCauliff}, S., {et~al.} 2016, \bibinfo{title}{{The TESS science processing operations center},} in Society of Photo-Optical Instrumentation Engineers (SPIE) Conference Series, Vol. 9913, Software and Cyberinfrastructure for Astronomy IV, ed. G.~{Chiozzi} \& J.~C. {Guzman}, 99133E, \dodoi{10.1117/12.2233418}

\bibitem[{D. {Jewitt} {et~al.}(2025){Jewitt}, {Hui}, {Mutchler}, {Kim}, \& {Agarwal}}]{2025ApJ...990L...2J}
{Jewitt}, D., {Hui}, M.-T., {Mutchler}, M., {Kim}, Y., \& {Agarwal}, J. 2025, \bibinfo{title}{{Hubble Space Telescope Observations of the Interstellar Interloper 3I/ATLAS},} \apjl, 990, L2, \dodoi{10.3847/2041-8213/adf8d8}

\bibitem[{D. {Jewitt} \& D.~Z. {Seligman}(2023){Jewitt} \& {Seligman}}]{2023ARA&A..61..197J}
{Jewitt}, D., \& {Seligman}, D.~Z. 2023, \bibinfo{title}{{The Interstellar Interlopers},} \araa, 61, 197, \dodoi{10.1146/annurev-astro-071221-054221}

\bibitem[{T. {Kareta} {et~al.}(2025){Kareta}, {Champagne}, {McClure}, {Emery}, {Sharkey}, {Bauer}, {Connelly}, {Rayner}, {Thomas}, {Reddy}, \& {Firgard}}]{2025arXiv250712234K}
{Kareta}, T., {Champagne}, C., {McClure}, L., {et~al.} 2025, \bibinfo{title}{{Near-Discovery Observations of Interstellar Comet 3I/ATLAS with the NASA Infrared Telescope Facility},} arXiv e-prints, arXiv:2507.12234, \dodoi{10.48550/arXiv.2507.12234}

\bibitem[{ {Lightkurve Collaboration} {et~al.}(2018){Lightkurve Collaboration}, {Cardoso}, {Hedges}, {Gully-Santiago}, {Saunders}, {Cody}, {Barclay}, {Hall}, {Sagear}, {Turtelboom}, {Zhang}, {Tzanidakis}, {Mighell}, {Coughlin}, {Bell}, {Berta-Thompson}, {Williams}, {Dotson}, \& {Barentsen}}]{2018ascl.soft12013L}
{Lightkurve Collaboration}, {Cardoso}, J.~V.~d.~M., {Hedges}, C., {et~al.} 2018, {Lightkurve: Kepler and TESS time series analysis in Python},, Astrophysics Source Code Library \doeprint{1812.013}

\bibitem[{T. {Nguyen} {et~al.}(2024){Nguyen}, {Woods}, {Ruprecht}, \& {Birge}}]{2024AJ....167..113N}
{Nguyen}, T., {Woods}, D.~F., {Ruprecht}, J., \& {Birge}, J. 2024, \bibinfo{title}{{Efficient Search and Detection of Faint Moving Objects in Image Data},} \aj, 167, 113, \dodoi{10.3847/1538-3881/ad20e0}

\bibitem[{A. {P{\'a}l} {et~al.}(2020){P{\'a}l}, {Szak{\'a}ts}, {Kiss}, {B{\'o}di}, {Bogn{\'a}r}, {Kalup}, {Kiss}, {Marton}, {Moln{\'a}r}, {Plachy}, {S{\'a}rneczky}, {Szab{\'o}}, \& {Szab{\'o}}}]{2020ApJS..247...26P}
{P{\'a}l}, A., {Szak{\'a}ts}, R., {Kiss}, C., {et~al.} 2020, \bibinfo{title}{{Solar System Objects Observed with TESS{\textemdash}First Data Release: Bright Main-belt and Trojan Asteroids from the Southern Survey},} \apjs, 247, 26, \dodoi{10.3847/1538-4365/ab64f0}

\bibitem[{L. {Pipeline} {et~al.}(2025){Pipeline}, {Ofek}, {Auto}, {Ye}, {Collaboration}, {Cattoi}, {Mancuso}, {Cioni}, {Parente}, {Agnetti}, {Sannino}, {Mercanti}, {Gentile}, {Hidas}, {Minev}, {Mutafov}, {Kostov}, {Erasmus}, {Robinson}, {Fitzsimmons}, {Denneau}, {Tonry}, {Weiland}, {Siverd}, {:unkn}, {Ivanova}, {Lysenko}, {Ivanov}, {Roshchupko}, {Yakovenko}, {Barcov}, {Ivanov}, {Ozhered}, {Kurbatov}, {Archibasov}, {Nedelcu}, {Sonka}, {Husar}, {Junius}, {Yoshimoto}, {Reina}, {Camarasa}, {Hornoch}, {Pravec}, {Micheli}, {Prystavski}, {Holt}, {Chatelain}, {Greenstreet}, {Gomez}, {Lister}, {Ventre}, {Betoret}, {Naves}, {Campas}, {Mazzanti}, {Tombelli}, {Bernardi}, {Interrante}, {Squilloni}, {Cheli}, {Loncao}, {Lombardo}, {Iozzi}, {Dupouy}, {Laborde}, {Diepvens}, {Bryssinck}, {Barreto}, {Martinez}, {Andre}, {Bosch}, {Jacques}, {Devogele}, {Ieva}, {Cellino}, {Kwon}, {Bendjoya}, {Gray}, {Bagnulo}, {Kolokolova}, {Hainaut}, {Maury}, {Signoret}, {Mari}, {Parrott}, {Attard}, {Usher}, {Rhemann}, {Jaeger}, {Ruocco},
  {Haeusler}, {Koch}, {Serra-Ricart}, {Alarcon}, {Romanov}, {De Queiroz}, {Maikner}, \& {Badcock}}]{2025MPEC....O...20P}
{Pipeline}, L., {Ofek}, E., {Auto}, L., {et~al.} 2025, \bibinfo{title}{{Comet 3I/ATLAS},} Minor Planet Electronic Circulars, 2025-O20, \dodoi{10.48377/MPEC/2025-O20}

\bibitem[{G.~R. {Ricker} {et~al.}(2015){Ricker}, {Winn}, {Vanderspek}, {Latham}, {Bakos}, {Bean}, {Berta-Thompson}, {Brown}, {Buchhave}, {Butler}, {Butler}, {Chaplin}, {Charbonneau}, {Christensen-Dalsgaard}, {Clampin}, {Deming}, {Doty}, {De Lee}, {Dressing}, {Dunham}, {Endl}, {Fressin}, {Ge}, {Henning}, {Holman}, {Howard}, {Ida}, {Jenkins}, {Jernigan}, {Johnson}, {Kaltenegger}, {Kawai}, {Kjeldsen}, {Laughlin}, {Levine}, {Lin}, {Lissauer}, {MacQueen}, {Marcy}, {McCullough}, {Morton}, {Narita}, {Paegert}, {Palle}, {Pepe}, {Pepper}, {Quirrenbach}, {Rinehart}, {Sasselov}, {Sato}, {Seager}, {Sozzetti}, {Stassun}, {Sullivan}, {Szentgyorgyi}, {Torres}, {Udry}, \& {Villasenor}}]{2015JATIS...1a4003R}
{Ricker}, G.~R., {Winn}, J.~N., {Vanderspek}, R., {et~al.} 2015, \bibinfo{title}{{Transiting Exoplanet Survey Satellite (TESS)},} Journal of Astronomical Telescopes, Instruments, and Systems, 1, 014003, \dodoi{10.1117/1.JATIS.1.1.014003}

\bibitem[{T. {Santana-Ros} {et~al.}(2025){Santana-Ros}, {Ivanova}, {Mykhailova}, {Erasmus}, {Kami{\'n}ski}, {Oszkiewicz}, {Kwiatkowski}, {Hus{\'a}rik}, {Ngwane}, \& {Penttil{\"a}}}]{2025arXiv250800808S}
{Santana-Ros}, T., {Ivanova}, O., {Mykhailova}, S., {et~al.} 2025, \bibinfo{title}{{Temporal Evolution of the Third Interstellar Comet 3I/ATLAS: Spin, Color, Spectra and Dust Activity},} arXiv e-prints, arXiv:2508.00808, \dodoi{10.48550/arXiv.2508.00808}

\bibitem[{D.~Z. {Seligman} {et~al.}(2025){Seligman}, {Micheli}, {Farnocchia}, {Denneau}, {Noonan}, {Hsieh}, {Santana-Ros}, {Tonry}, {Auchettl}, {Conversi}, {Devog{\`e}le}, {Faggioli}, {Feinstein}, {Fenucci}, {Ferrais}, {Frincke}, {Hainaut}, {Hart}, {Hoffman}, {Holt}, {Hoogendam}, {Huber}, {Jehin}, {Kareta}, {Keane}, {Kelley}, {Lister}, {Mandt}, {Mar{\v{c}}eta}, {Meech}, {Amine Miftah}, {Morgan}, {Oca{\~n}a}, {Pe{\~n}a-Asensio}, {Shappee}, {Siverd}, {Taylor}, {Tucker}, {Wainscoat}, {Weryk}, {Wray}, {Yaginuma}, {Yang}, {Ye}, \& {Zhang}}]{2025arXiv250702757S}
{Seligman}, D.~Z., {Micheli}, M., {Farnocchia}, D., {et~al.} 2025, \bibinfo{title}{{Discovery and Preliminary Characterization of a Third Interstellar Object: 3I/ATLAS},} arXiv e-prints, arXiv:2507.02757, \dodoi{10.48550/arXiv.2507.02757}

\bibitem[{ {STScI}(2022){STScI}}]{mast_tess_ffi}
{STScI}. 2022, TESS Calibrated Full Frame Images: All Sectors, STScI/MAST, \dodoi{10.17909/0CP4-2J79}

\bibitem[{N. {Tak{\'a}cs} {et~al.}(2025){Tak{\'a}cs}, {Kiss}, {Szak{\'a}ts}, \& {P{\'a}l}}]{2025PASP..137d4401T}
{Tak{\'a}cs}, N., {Kiss}, C., {Szak{\'a}ts}, R., \& {P{\'a}l}, A. 2025, \bibinfo{title}{{Solar System Objects Observed with TESS{\textemdash}Early Data Release 2: I. Spin-shape Recovery Potential of Multi-epoch TESS Observations},} \pasp, 137, 044401, \dodoi{10.1088/1538-3873/adc0c1}

\bibitem[{J. {Tonry} {et~al.}(2025){Tonry}, {Denneau}, {Alarcon}, {Clocchiatti}, {Erasmus}, {Fitzsimmons}, {Licandro}, {Meech}, {Siverd}, \& {Weiland}}]{2025arXiv250905562T}
{Tonry}, J., {Denneau}, L., {Alarcon}, M., {et~al.} 2025, \bibinfo{title}{{ATLAS Photometry of Interstellar Object 3I/ATLAS},} arXiv e-prints, arXiv:2509.05562, \dodoi{10.48550/arXiv.2509.05562}

\bibitem[{J.~L. {Tonry} {et~al.}(2018){Tonry}, {Denneau}, {Heinze}, {Stalder}, {Smith}, {Smartt}, {Stubbs}, {Weiland}, \& {Rest}}]{2018PASP..130f4505T}
{Tonry}, J.~L., {Denneau}, L., {Heinze}, A.~N., {et~al.} 2018, \bibinfo{title}{{ATLAS: A High-cadence All-sky Survey System},} \pasp, 130, 064505, \dodoi{10.1088/1538-3873/aabadf}

\bibitem[{A. Tuson {et~al.}(2025)Tuson, Martínez-Palomera, \& Hedges}]{tuson_2025_16332750}
Tuson, A., Martínez-Palomera, J., \& Hedges, C. 2025, altuson/tess-asteroids: v1.2.6, v1.2.6 Zenodo, \dodoi{10.5281/zenodo.16332750}

\bibitem[{G.~V. {Williams} {et~al.}(2017){Williams}, {Sato}, {Sarneczky}, {Wainscoat}, {Woodworth}, \& {Meech}}]{2017CBET.4450....1W}
{Williams}, G.~V., {Sato}, H., {Sarneczky}, K., {et~al.} 2017, \bibinfo{title}{{Minor Planets 2017 SN\_33 and 2017 U1},} Central Bureau Electronic Telegrams, 4450, 1

\bibitem[{D.~F. {Woods} {et~al.}(2021){Woods}, {Ruprecht}, {Kotson}, {Main}, {Evans}, {Varey}, {Vaillancourt}, {Viggh}, {Brown}, \& {P{\'a}l}}]{2021PASP..133a4503W}
{Woods}, D.~F., {Ruprecht}, J.~D., {Kotson}, M.~C., {et~al.} 2021, \bibinfo{title}{{Asteroid Observations from the Transiting Exoplanet Survey Satellite: Detection Processing Pipeline and Results from Primary Mission Data},} \pasp, 133, 014503, \dodoi{10.1088/1538-3873/abc761}

\bibitem[{B. {Yang} {et~al.}(2025){Yang}, {Meech}, {Connelley}, \& {Keane}}]{2025arXiv250714916Y}
{Yang}, B., {Meech}, K.~J., {Connelley}, M., \& {Keane}, J.~V. 2025, \bibinfo{title}{{Spectroscopic Characterization of Interstellar Object 3I/ATLAS: Water Ice in the Coma},} arXiv e-prints, arXiv:2507.14916, \dodoi{10.48550/arXiv.2507.14916}

\bibitem[{Q. {Ye} {et~al.}(2025){Ye}, {Kelley}, {Hsieh}, {Bellm}, {Chen}, {Dekany}, {Drake}, {Groom}, {Helou}, {Kulkarni}, {Prince}, \& {Riddle}}]{2025arXiv250908792Y}
{Ye}, Q., {Kelley}, M. S.~P., {Hsieh}, H.~H., {et~al.} 2025, \bibinfo{title}{{Prediscovery Activity of New Interstellar Object 3I/ATLAS: A Dynamically-Old Comet?},} arXiv e-prints, arXiv:2509.08792, \dodoi{10.48550/arXiv.2509.08792}

\end{thebibliography}
\bibliographystyle{aasjournalv7}

\end{document}